\documentstyle[eqsecnum,prd,aps,epsf,multicol]{revtex}
\newcommand{\para}{|\!|}
\newcommand{\sgn}{\mbox{sgn}}
\begin{document}
%\begin{titlepage}
\draft
\title{
  Does a Nambu-Goto wall emit gravitational waves?\\
  \smallskip 
  {\normalsize -- Cylindrical Nambu-Goto wall as an example of
    gravitating non-spherical walls --} 
  }
\author{Kouji Nakamura\footnote{E-mail:kouchan@th.nao.ac.jp} }
\address{Division of Theoretical Astrophysics, National
Astronomical Observatory, \\ Mitaka, Tokyo 181-8588, Japan}
\date{\today}
\maketitle

\begin{abstract}
Gravitational field of a cylindrical Nambu-Goto wall in the vacuum
spacetime is considered in order to clarify the interaction between
Nambu-Goto membranes and gravitational waves. 
If one neglects the emission of gravitational waves by the wall
motion, the spacetime becomes singular. 
It is also shown that the emission of gravitational waves does occur
by the motion of the cylindrical wall if the initial data is
singularity free. 
The energy loss rate due to radiation of gravitational waves agrees
with that estimated from the test wall motion and the quadrupole
formula for the gravitational wave emission. 
This is quite different from the oscillatory behavior of gravitating
Nambu-Goto membranes: the presence of gravity induces the wall to lose
its dynamical degree of freedom.
\end{abstract}
%\end{titlepage}

\begin{multicols}{2}
  
%%%%%%%%%%%%%%%%%%%%%%%%%%%%%%%%%%%%%%%%%%%%%%%%%%%%%%%%%%%%%%%%%
\section{Introduction}
\label{sec:intro}
%%%%%%%%%%%%%%%%%%%%%%%%%%%%%%%%%%%%%%%%%%%%%%%%%%%%%%%%%%%%%%%%%

%*****************************************************************

Gravitational field and dynamics of extended objects (membranes) are
familiar topics in recent physics. 
Historically, gravity of membranes is investigated in the context of
topological defects formed during phase transition in the early
universe\cite{Vilenkin_Shellard,Kolb-Turner}. 
Among these extended objects, domain walls are the simplest type of
defects which arose when the phase transition has occurred by the
breakdown of a discrete symmetry. 
However, the theoretical prediction using standard scaling defect
models are in conflict with observations of the Cosmic Background
Explorer satellite\cite{Albrecht-Battye-Robinson}. 
Further, stable domain wall networks are not allowed in the early
universe because they overdominate the radiation energy
density\cite{Vilenkin_Shellard,Kolb-Turner}. 
Despite of these discouraging results, topological defects have been
remained intrinsically interesting topics to study extended objects. 
Unstable defects formed in the early universe might radiate
gravitational waves by their rapid oscillation\cite{Vachaspati} and
then the detection of relic gravitational waves from these extended
objects is considered to be one of the evidences for the phase
transition in the universe. 
Moreover, as seen in the recent proposal of the theoretical models with
the large extradimension\cite{Randall-Sundrum}, so called brane world
scenario, domain walls are considered as a realization of our universe
in higher dimensions.

%*****************************************************************

Within the general relativity, topological defects are unusual and
are expected to play a role of an interesting source of gravity. 
In the simplest case, domain walls are idealized by the
infinitesimally thin Nambu-Goto membranes. 
If the self-gravity of membranes is ignored (test membrane case), the
Nambu-Goto action admits oscillatory solutions which radiate
gravitational waves as mentioned above.  
However, by taking into account the self-gravity of the Nambu-Goto
wall, it is shown that self-gravitating walls coupled to gravitational
wave behave in a quite different manner.
The dynamical degrees of freedom concerning the perturbative
oscillations around spherical walls is given by that of gravitational
waves, and self-gravitating spherical walls do not oscillate
spontaneously contrary to the case of test walls\cite{KoIshiFuji}.
The essentially same conclusion is also obtained in the case of an
infinite Nambu-Goto string\cite{kouchan-string}. 
These results insist that the motion of gravitating membranes is quite
different from the motion of the test membranes even if the total
energy of the wall is very small. 
This difference between test membranes and self-gravitating membranes
is crucial when one estimates the energy of the gravitational waves
from these extended objects.

%*****************************************************************

In this paper, we investigate the gravitational wave emission due to
the motion of a Nambu-Goto wall. 
Particularly, we consider the cylindrically symmetric domain wall as a
toy model of highly elongated gravitating walls. 
The results that the gravitating Nambu-Goto membrane has no dynamical
degree of freedom of are based on the perturbation around the spacetime 
without gravitational wave. 
We expect that the behavior of highly distorted Nambu-Goto
walls is different. 
The cylindrically symmetric spacetime considered here is described by
the metric of the {\it Weyl canonical form}. 
In this metric, there are cylindrically symmetric wave modes, which is
called {\it Einstein-Rosen wave} (ER wave), and a domain wall's motion
excites ER wave. 
We have to treat the coupled system of ER wave and the cylindrically
symmetric domain wall, which is governed by the Einstein equation and
the junction condition\cite{Israel} on the wall.

%*****************************************************************

We first show that all spacetimes containing a cylindrical wall
become singular if we neglect ER wave emission. 
This implies that, in the absence of singularities, ER wave emission
necessarily occurs by the motion of the cylindrical wall.
Next, we consider the momentarily static ``regular'' initial data for
the spacetime containing a cylindrical domain wall and their
infinitesimal time evolution. 
We show that the monotonically collapsing motion of the gravitating
cylindrical wall does emit gravitational waves and that the estimate
of the energy of gravitational waves using the quadrupole formula is
correct if the wall energy density is sufficiently small. 
This is the quite reasonable result expected from the behavior of the
Nambu-Goto membrane in the absence of gravity.
However, we have to notice that this result has an apparent 
discrepancy with the results in Ref.\cite{KoIshiFuji,kouchan-string}.

%*****************************************************************

The organization of this paper is as follows. 
In the next section, we briefly review the symmetry reduction of
spacetime into the cylindrically symmetric one. 
We also derive the basic equations that govern the motion of the
domain wall.
In Sec.III, we consider the possible solutions of a self-gravitating
domain wall without gravitational wave emission. 
In Sec.IV, we set up momentarily static and radiation free regular
initial configurations and investigate the time evolution to see the
gravitational wave emission due to the wall motion. 
Finally, Sec.V is devoted to the summary and discussions about the
discrepancy with the conclusion in
Ref.\cite{KoIshiFuji,kouchan-string}.

%*****************************************************************

Throughout this paper, we use the unit such that the light velocity
$c=1$ and Newton's gravitational constant is denoted by $G$. 
The signature of Lorentzian metrics is chosen to be $(-,+,+,+)$.

%*****************************************************************

%%%%%% Section 2 ``Cylindrical Spacetime and Domain Wall'' %%%%%%%%%%%%
\section{Cylindrical Spacetime with a Domain Wall}
\label{sec:general}
%%%%%%%%%%%%%%%%%%%%%%%%%%%%%%%%%%%%%%%%%%%%%%%%%%%%%%%%%%%%%%%%%%%%%%%

%*****************************************************************

We consider the spacetime with a cylindrical domain wall using the
thin wall approximation. 
The whole spacetime $({\cal M},g_{ab})$ contains two vacuum regions
$({\cal M}_{+}, g_{ab+})$ and $({\cal M}_{-}, g_{ab-})$. 
Each region (${\cal M}_{+}$ or ${\cal M}_{-}$) has a timelike boundary
$\Sigma_{\pm}$, respectively, which should be identified so that 
$\Sigma:=\Sigma_{+}=\Sigma_{-}$ using Israel's junction
condition\cite{Israel}. 
The whole spacetime is ${\cal M}={\cal M}_{-}\cup\Sigma\cup{\cal M}_{+}$. 
The timelike submanifold $\Sigma$ is the world volume of the domain wall.

%*****************************************************************

Since we consider the domain wall with cylindrical symmetry, it is
natural to consider the vacuum region ${\cal M}_{\pm}$ also have
cylindrical symmetry.
The cylindrically symmetric spacetime considered in this paper has two
commutable spacelike Killing vector fields $z^{a}$ and $\phi^{a}$ and
these are both hypersurface orthogonal.
The orbit of $z^{a}$ is $R^{1}$ and that of $\phi^{a}$ is $S^{1}$.
We introduce coordinate functions $z$ and $\phi$ by $z^{a}\nabla_{a}z
= 1$ and $\phi^{a}\nabla_{a}\phi = 1$, respectively.
$\phi$ is a periodic coordinate with the period $2\pi$.

%*****************************************************************

Using one of the vacuum Einstein equations 
$R^{z}_{z} + R^{\phi}_{\phi} = 0$ and assuming that the gradient of
$\sqrt{(\phi^{a}\phi_{a})(z^{a}z_{a})}$ is spacelike, the both metrics
on the spacetime ${\cal M}_{\pm}$ are reduced to the 
{\it Weyl canonical form}:   
\begin{equation}
  \label{Weyl-cano}
  ds^{2} = e^{2(\gamma - \psi)} (- dt^{2} + dr^{2}) + e^{2\psi} dz^{2} 
   + r^{2} e^{-2\psi}d\phi^{2},
\end{equation}
where $\psi$ and $\gamma$ depend on $t$ and $r$.
The existence of a thin wall does not contradict to the assumption
that the gradient of $\sqrt{(\phi^{a}\phi_{a})(z^{a}z_{a})}$ is
spacelike.
(See Appendix \ref{sec:closed_non_closed}.)
The coordinates $t$ and $r$ parameterize the two-dimensional orbit
space ${\cal N}$ and each point on $\cal N$ corresponds to a cylinder
of symmetry.
Further, we call the direction to which $t$ increases (decreases)
``future'' (``past'') direction.

%*************************************************************

The vacuum Einstein equations for the metric (\ref{Weyl-cano}) are
given by 
\begin{eqnarray}
  \label{ER-wave}
  && \partial_{t}^{2}\psi - \frac{1}{r}\partial_{r}\left(r\partial_{r}
  \psi\right) = 0, \\
  \label{gammar}
  && \partial_{r}\gamma = r \left((\partial_{t}\psi)^{2} +
  (\partial_{r}\psi)^{2}\right), \\
  \label{gammat}
  && \partial_{t}\gamma = 2 r (\partial_{t}\psi) (\partial_{r}\psi). 
\end{eqnarray}
Eq.(\ref{ER-wave}) is the wave equation and $\psi$ corresponds to the
plus mode of gravitational waves, so called {\it Einstein-Rosen wave}
(ER wave).

%*************************************************************
%%%%%%%%%%%%%%%%%%%%%%%%%%%%%%%%%%%%%%%%%%%%%%%%%%%%%%%%%%%%%%%%%%%%%%
%\newpage
%%%%%%%%%%%%%%%%%%%%%%%%%%%%%%%%%%%%%%%%%%%%%%%%%%%%%%%%%%%%%%%%%%%%%%% 
%*************************************************************

The timelike curve $\Sigma\cap{\cal N}$ is the trajectory of the
domain wall in ${\cal N}$.
The future directed unit tangent $u^{a}$ ($u^{a}u_{a} = -1$) of this
timelike curve is the 4-velocity of the domain wall and the proper
time $\tau$ of the domain wall is introduced by $u^{a}\nabla_{a}\tau=1$.
Since we consider the cylindrically symmetric domain wall with the
Killing vectors $z^{a}:=z^{a}_{\pm}$ and  $\phi^{a}:=\phi^{a}_{\pm}$
on $\Sigma_{\pm}$, the coordinate functions $z$ and $\phi$ on 
${\cal M}_{\pm}$ are extended to smooth functions on the whole spacetime
${\cal M}$.
Then the coordinate system $(\tau,z,\phi)$ on $\Sigma$ is naturally
induced.

%*************************************************************

The induced metrics on $\Sigma_{\pm}$ from $g_{ab\pm}$ are given by
\begin{eqnarray}
  \label{induced-pm-metric}
  h_{ab\pm} &=& - (d\tau)_{a}(d\tau)_{b} 
  + e^{2\Psi_{\pm}(\tau)}(dz)_{a}(dz)_{b} \nonumber\\
  && \quad\quad + R_{\pm}^{2}(\tau)
  e^{-2\Psi_{\pm}(\tau)}(d\phi)_{a}(d\phi)_{b},
\end{eqnarray}
where $\Psi_{\pm}(\tau)=\left.\psi\right|_{\Sigma_{\pm}}$ and
$R_{\pm}(\tau)=\left.r\right|_{\Sigma_{\pm}}$, respectively.
Since $\Sigma_{\pm}$ should be diffeomorphic to each other to identify
them, $h_{ab\pm}$ must satisfy $h_{ab}:=h_{ab+}=h_{ab-}$ in the above
coordinate system: 
\begin{equation}
  \label{explicit_intrinsic_junction}
  R(\tau) := R_{+} = R_{-}, \quad \Psi(\tau) := \Psi_{+} = \Psi_{-}.
\end{equation}

%*************************************************************

Further, the identification should be done so that $n_{a+}=n_{a-}$,
where $n_{a\pm}$ are unit normal to $\Sigma_{\pm}$ directed from
${\cal M}_{-}$ to ${\cal M}_{+}$.
The existence of the domain wall gives the finite discontinuity of the
extrinsic curvature $K^{a}_{b\pm}$ of $\Sigma_{\pm}$ in 
${\cal M}_{\pm}$, where  
$K^{a}_{b\pm} := h^{ac}h^{d}_{b}\nabla_{c}^{\pm}n_{d\pm}$, 
and $\nabla_{a}^{\pm}$ is the covariant derivatives associated with
the metric $g_{ab\pm}$, respectively. 
When the surface energy on $\Sigma$ is given by 
$S^{a}_{b} = \sigma h^{a}_{b}$, the Israel's junction conditions
becomes to $[K^{a}_{b}] = - \lambda h^{a}_{b}$\cite{notation-subst}.
$\lambda = 4\pi G \sigma>0$ is the surface tension of the wall which
is equal to the surface density.

%*************************************************************

In terms of the coordinate system in Eq.(\ref{Weyl-cano}), $u^{a}$ is
given by
\begin{eqnarray}
  \label{UrUt_def}
  && u^{a} = u^{t}_{\pm} \left(\frac{\partial}{\partial t}\right)^{a}_{\pm}
  + u^{r} \left(\frac{\partial}{\partial r}\right)^{a}_{\pm},\\
  \label{UrUt_comp}
  && u^{r} = \frac{dR}{d\tau}, \;\;
   u^{t}_{\pm} = \frac{dT_{\pm}}{d\tau} =
  \sqrt{\left(\frac{dR}{d\tau}\right)^2 + e^{-2(\Gamma_{\pm} -
      \Psi_{s})}}, 
\end{eqnarray}
where $\Gamma_{\pm}:=\left.\gamma(t,r)\right|_{\Sigma_{\pm}}$ and
$T_{\pm} := \left.t\right|_{\Sigma_{\pm}}$.
$u^{t}_{\pm}$ are positive because $u^{a}$ is future directed.
From the orthonormal condition $n^{a}_{\pm}n_{a\pm} = 1$ and
$u^{a}n_{a\pm} = 0$, $n^{a}_{\pm}$ are given by  
\begin{equation}
  \label{NrNt_comp}
  n^{a}_{\pm} = \epsilon_{\pm} \left(
    u^{r} \left(\frac{\partial}{\partial t_{\pm}}\right)^{a}_{\pm} 
    + 
    u^{t}_{\pm} \left(\frac{\partial}{\partial r}\right)^{a}_{\pm} 
  \right),
\end{equation}
where $\epsilon_{\pm} = \sgn(n^{a}_{\pm}\partial_{a}r)$.
Using Eqs.(\ref{UrUt_def})-(\ref{NrNt_comp}), the all components of
the junction condition for the extrinsic curvature of $\Sigma$ are
given by 
\begin{eqnarray}
  \label{junction-tautau}
  && \left(\frac{d^2 R}{d\tau^2} - \frac{d R}{d\tau} D_{\parallel}\psi
    + R\left(D_{\parallel}\psi\right)^2\right)
  \left[\frac{1}{D_{\perp}r}\right] \nonumber\\
  && \quad\quad\quad + 
  R\left[\frac{\left(D_{\perp}\psi\right)^2}{D_{\perp}r}\right] = -
  2\lambda, \\    
  \label{junction-psi}
  && [D_{\perp}\psi] = - \lambda, \\
  \label{junction-r}
  && [D_{\perp}r] = - 2 \lambda R.
\end{eqnarray}
where $D_{\parallel}A = u^{a}\nabla_{a}A$, $(D_{\perp}A)_{\pm} =
(n^{a}\nabla_{a}A)_{\pm}$ for an arbitrary function $A$ in the
neighborhood of $\Sigma$ and the subscript $\pm$ shows the functions
evaluated on $\Sigma_{\pm}$, respectively.
Further, we note $[D_{\para}\psi] = 0 = [D_{\para}r]$.

%*************************************************************

Eq.(\ref{junction-tautau}) is the equation of motion for the
cylindrical wall $\Sigma$ and Eq.(\ref{junction-psi}) is the boundary
condition for ER wave $\psi$ on $\Sigma$. 
Eq.(\ref{junction-r}) is also rewritten by  
\begin{eqnarray}
    && \left(\frac{dR}{d\tau}\right)^2 - (\lambda R)^{2} 
    - \frac{1}{4\lambda R} \left(e^{-2\Gamma_{+}} -
      e^{-2\Gamma_{-}} \right)^{2} \nonumber\\
    && \quad \quad  
  + \frac{1}{2} \left(e^{-2\Gamma_{+}} + e^{-2\Gamma_{-}}
  \right) e^{\Psi} = 0
  \label{general_eq_of_motion}
\end{eqnarray}
by virtue of Eq.(\ref{NrNt_comp}).
We note that Eq.(\ref{junction-tautau}) is derived from
Eqs.(\ref{junction-psi}) and (\ref{junction-r}) by using
Eqs.(\ref{ER-wave})-(\ref{gammat}) and Eq.(\ref{UrUt_comp}).

%*************************************************************

We explicitly see that the behavior of the ER wave at the boundary
($\Psi$ and $\Gamma_{\pm}$) affects the wall motion by
Eq.(\ref{general_eq_of_motion}) and the wall motion affects $\psi$
and $\gamma$ by their boundary conditions Eq.(\ref{junction-psi}).
Thus, this is the radiation reaction problem.
To clarify the dynamics of a gravitating cylindrical domain wall, we
have to solve the equations Eqs.(\ref{ER-wave})-(\ref{gammat}) with
the boundary conditions Eq.(\ref{junction-psi}) and
Eq.(\ref{junction-r}) on $\Sigma$ simultaneously, in general. 
In this paper, we show that the cylindrical domain wall must emit
gravitational waves using these equations and boundary conditions.

%%%%%%%%%%%%%%%%%%%%%%%%%%%%%%%%%%%%%%%%%%%%%%%%%%%%%%%%%%%%%%%%%%%%%%%
%\newpage
%%%%%%%%%%%%%%%%%%%%%%%%%%%%%%%%%%%%%%%%%%%%%%%%%%%%%%%%%%%%%%%%%%%%%%% 
%%%%%%%%%%%%%%%%%%%%%%% ``Waveless solutions'' %%%%%%%%%%%%%%%%%%%%%%%% 
\section{self-gravitating domain wall without gravitational wave emission} 
\label{sec:waveless} 
%%%%%%%%%%%%%%%%%%%%%%%%%%%%%%%%%%%%%%%%%%%%%%%%%%%%%%%%%%%%%%%%%%%%%%% 

%*************************************************************

In this section, we consider the self-gravitating domain wall
spacetime without gravitational wave emission and show that all
solutions contain singular axes.

%*************************************************************

Since ${\cal M}_{\pm}$ considered here is static, we assume that 
\begin{equation}
  \partial_{t}^{2}\psi = \partial_{t}\psi = 0.
\end{equation}
The static solution to Eqs.(\ref{ER-wave})-(\ref{gammat}) is
\begin{equation}
  \label{waveless_solutions}
  \psi = - \kappa \ln \left(\frac{r}{R_{0}}\right), \quad 
  \gamma = \gamma_{0} + \kappa^{2} \ln\left(\frac{r}{R_{0}}\right).
\end{equation}
where $R_{0}$, $\gamma_{0}$ and $\kappa$ are constants and the line
element is given by
\begin{eqnarray}
  ds^{2} &=& 
  e^{2\gamma_{0}}\left(\frac{r}{R_{0}}\right)^{2(\kappa^{2}+\kappa)} 
  (- dt^{2} + dr^{2}) \nonumber\\
  && \quad\quad +  \left(\frac{r}{R_{0}}\right)^{- 2\kappa} dz^{2}    
   + r^{2} \left(\frac{r}{R_{0}}\right)^{2\kappa}d\phi^{2}. 
  \label{waveless_metric}
\end{eqnarray}
This is wel-known as the Levi-Civita metric\cite{Exact-solutions}.

%*************************************************************

The circumferential radius of the symmetric cylinder $dt=dr=0$,
\begin{equation}
  \label{circumference_phiphi}
  \tilde{r}(r) := \frac{1}{2\pi} \int_{0}^{2\pi}d\phi \sqrt{g_{\phi\phi}},
\end{equation}
tells us the axis of the cylindrical symmetry.
When $\kappa+1>0$, $\tilde{r}(r)$ is a monotonically increasing
function of $r$ and vanishes at $r=0$. This means that $r=0$ is the
axis of symmetry in this case.
On the other hand, $\tilde{r}(r)$ monotonically decreases and vanishes
at $r=\infty$ when $\kappa+1<0$. 
Then, $r=\infty$ is the axis of cylindrical symmetry in the case of
$\kappa+1<0$.

%*************************************************************

The square of the Riemann curvature for the metric
(\ref{waveless_metric}) is given by 
\begin{eqnarray}
  I &:=& R_{abcd}R^{abcd} \nonumber\\
  &=& \frac{16\kappa^2 (1 + \kappa)^2 (1 + \kappa + \kappa^2)}
  {R_{0}^{4}e^{4\gamma_{0}}}
  \left(\frac{R_{0}}{r}\right)^{4(\kappa^2+\kappa+1)}.  
  \label{riemann_polynomial}
\end{eqnarray}
We note that the curvature of the spacetime approaches to zero as $I
\propto r^{-3-(2\kappa+1)^{2}} \rightarrow 0$ only when
$r\rightarrow\infty$. 
On the other hand, when $r\rightarrow 0$, $I$ diverges, except the
cases $\kappa = 0$ or $-1$. 
When $\kappa = 0$ or $-1$, the metric (\ref{waveless_metric}) is
locally flat.

%*************************************************************

Here, we consider the construction of the whole spacetime 
${\cal M}={\cal M}_{+}\cup\Sigma\cup{\cal M}_{-}$ with the metrics
(\ref{waveless_metric}) on ${\cal M}_{\pm}$.
We denote $\gamma_{0}$ and $\kappa$ on each ${\cal M}_{\pm}$ by
$\gamma_{0\pm}$ and $\kappa_{\pm}$, respectively.
The junction conditions Eqs.(\ref{explicit_intrinsic_junction}) are
given by 
\begin{equation}
  \label{waveless_intrinsic_condition}
  \ln\left(\frac{R}{R_{0}}\right)^{\kappa_{+}} =
  \ln\left(\frac{R}{R_{0}}\right)^{\kappa_{-}}, 
\end{equation}
and the conditions (\ref{junction-psi}) and (\ref{junction-r}) become
to 
\begin{eqnarray}
  \label{waveless-junction-1}
  && \epsilon_{+} \kappa_{+} u^{t}_{+} 
    - \epsilon_{-} \kappa_{-} u^{t}_{-} = \lambda R \\
  \label{waveless-junction-2}
  && \epsilon_{+} u^{t}_{+} - \epsilon_{-} u^{t}_{-} = - 2 \lambda R.
\end{eqnarray}
To evaluate
Eqs.(\ref{waveless_intrinsic_condition})-(\ref{waveless-junction-2}), 
and (\ref{junction-tautau}), we consider
two cases, $\kappa_{+}\neq\kappa_{-}$ and $\kappa_{+}=\kappa_{-}$,
separately.

%%%%%%%%%%%%%%%%%%%%%%%%%%%%%%%%%%%%%%%%%%%%%%%%%%%%%%%%%%%%%%%%%%%%%%%
%\newpage
%%%%%%%%%%%%%%%%%%%%%%%%%%%%%%%%%%%%%%%%%%%%%%%%%%%%%%%%%%%%%%%%%%%%%%%
%%%%%%%%%%%%%%%%%%%%%%%%%%%%%%%%%%%%%%%%%%%%%%%%%%%%%%%%%%%%%%%%%%%%%%% 
\subsection{$\kappa_{+}\neq\kappa_{-}$ case}
\label{sec:kappa+neqkappa-case}
%%%%%%%%%%%%%%%%%%%%%%%%%%%%%%%%%%%%%%%%%%%%%%%%%%%%%%%%%%%%%%%%%%%%%%% 

In this case, Eq.(\ref{waveless_intrinsic_condition}) must hold for
arbitrary $\tau$, i.e.,
\begin{equation}
  \label{static_wall_condition}
  R(\tau) = R_{0}, \quad \frac{dR}{d\tau} = 0.
\end{equation}
These mean the domain wall should stay at $r=R_{0}$. 
Then the components of the 4-velocity (\ref{UrUt_comp}) are given by
\begin{equation}
  \label{ur-utpm}
  u^{r} = 0, \quad u^{t}_{\pm} = e^{-\gamma_{0\pm}}.
\end{equation}
The conditions (\ref{waveless-junction-1}) and
(\ref{waveless-junction-2}) are given by  
\begin{equation}
  \label{kpkm}
  \epsilon_{\pm} u^{t}_{\pm} = \frac{1 + 2\kappa_{\mp}}{\kappa_{+} -
    \kappa_{-}}  \lambda R_{0}, 
\end{equation}
which shows that $\epsilon_{\pm}$ are determined by
\begin{equation}
  \label{epsilon_determination}
  \epsilon_{\pm} = \sgn\left(\frac{1 + 2\kappa_{\mp}}{\kappa_{+} -
      \kappa_{-}}\right).
\end{equation}

%*************************************************************

Eq.(\ref{junction-tautau}) gives the acceleration of the wall:
\begin{eqnarray}
  \frac{d^{2}R}{d\tau^{2}} 
   &=& - \frac{(1+2\kappa_{+})(1+2\kappa_{-})}{2(\kappa_{+}-\kappa_{-})^{2}} 
   \times
   \nonumber\\
  && \quad
  (\kappa_{+} + \kappa_{-} + 2 \kappa_{+}\kappa_{-} + 2) \lambda^{2}R_{0}. 
  \label{acceleration_kappa-neqkappa+}
\end{eqnarray}
Since Eqs.(\ref{static_wall_condition}) should hold for arbitrary
 $\tau$, $(d^{2}R/d\tau^{2})=0$ for arbitrary $\tau$.
We easily see that $\kappa_{\pm}\neq 1/2$ from Eqs.(\ref{kpkm}) and
 $\kappa_{-}\neq\kappa_{+}$.
Then $\kappa_{\pm}$ should satisfy the relation
\begin{equation}
  \label{Tomita-relation}
  \kappa_{+} + \kappa_{-} + 2 \kappa_{+}\kappa_{-} + 2 = 0.
\end{equation}
This is the condition obtained by Tomita\cite{Tomita}.

%*************************************************************

From the relation (\ref{Tomita-relation}) and
Eq.(\ref{epsilon_determination}), we obtain
$(\epsilon_{-},\epsilon_{+})=(+,-)$, which means that the coordinate
function $r$ is maximum at $\Sigma$ and both ${\cal M}_{\pm}$ include
the axis $r=0$. 
Eq.(\ref{Tomita-relation}) also forbids the cases that both of
$\kappa_{\pm}$ are $0$ or $-1$. 
Then, one of the axis $r=0$ has to be singular.

%%%%%%%%%%%%%%%%%%%%%%%%%%%%%%%%%%%%%%%%%%%%%%%%%%%%%%%%%%%%%%%%%%%%%%
%\newpage
%%%%%%%%%%%%%%%%%%%%%%%%%%%%%%%%%%%%%%%%%%%%%%%%%%%%%%%%%%%%%%%%%%%%%%% 
\subsection{$\kappa_{+}=\kappa_{-}$ case}
\label{sec:kappa+=kappa-case}
%%%%%%%%%%%%%%%%%%%%%%%%%%%%%%%%%%%%%%%%%%%%%%%%%%%%%%%%%%%%%%%%%%%%%%% 

In this case, we obtain $\kappa_{\pm}=-1/2$ from
Eqs.(\ref{waveless-junction-1}) and (\ref{waveless-junction-2}). 
The condition Eq.(\ref{waveless_intrinsic_condition}) is trivially
satisfied. 
Then, the domain wall may move on ${\cal N}$.
The equation (\ref{general_eq_of_motion}) of the wall motion is given by  
\begin{eqnarray}
  \label{eq_of_motion_kappa+=kappa-}
  && \left(\frac{dR}{d\tau}\right)^{2} + V(R) = 0,\\
  \label{potential_kappa+=kappa-}
  && V(R) = - (\lambda R)^{2} - \frac{1}{(4\lambda)^{2}RR_{0}} 
  \left(e^{-2\gamma_{0+}}-e^{-2\gamma_{0-}}\right) \nonumber\\
  && \quad \quad + \frac{1}{2} \left(e^{-2\gamma_{0-}} 
    + e^{-2\gamma_{0+}}\right)
  \left(\frac{R}{R_{0}}\right)^{1/2}.
\end{eqnarray}
Actually, Eq.(\ref{eq_of_motion_kappa+=kappa-}) has the solutions of a
moving domain wall without gravitational waves\cite{Ipser-Sikivie-comment}.
From the energy condition, $\lambda > 0$, and the condition
(\ref{waveless-junction-2}), the case
$(\epsilon_{-},\epsilon_{+})=(-,+)$ is rejected.
Then, ${\cal M}$ must have the axis $r=0$ in itself.
Since $\kappa_{\pm}=-1/2$ (neither $0$ nor $-1$), the axis is
singular.

%%%%%%%%%%%%%%%%%%%%%%%%%%%%%%%%%%%%%%%%%%%%%%%%%%%%%%%%%%%%%%%%%%%%%%
%\newpage
%%%%%% Section 4 ``Momentarily Static Initial Configuration'' %%%%%%%%% 
\section{Einstein-Rosen Wave Emission from cylindrical domain wall} 
\label{sec:radiation}
%%%%%%%%%%%%%%%%%%%%%%%%%%%%%%%%%%%%%%%%%%%%%%%%%%%%%%%%%%%%%%%%%%%%%%%
%%%%%%%%%%%%%%%%%%%%%%%%%%%%%%%%%%%%%%%%%%%%%%%%%%%%%%%%%%%%%%%%%%%%%%%

In the previous section, we have seen the all spacetimes with a
self-gravitating cylindrical domain wall without gravitational waves
have to include singular axes in itself.
This implies that we should take into account the emission of
gravitational waves if we require the regularity of the spacetime and
the emission of gravitational waves necessarily occurs.
To see this, we consider the infinitesimal time evolution from a
``{\it regular}'' initial configuration.
In this paper, we regard that the initial surface is {\it regular} if
the initial surface does not include scalar polynomial singularities
or deficit angles except the delta-function matter distribution of the
wall.
We concentrate on the momentarily static and radiation free initial
configuration, and its infinitesimal time evolution, for simplicity.
Then, we see that ER wave emission does occur due to the motion of the
wall.

%*************************************************************

The total system of a self-gravitating domain wall in the
cylindrically symmetric spacetime is governed by the Einstein
equations (\ref{ER-wave})-(\ref{gammat}), the junction conditions for
the intrinsic metric (\ref{explicit_intrinsic_junction}), and those for
the extrinsic curvature (\ref{junction-tautau})-(\ref{junction-r}).
Among them, Eqs.(\ref{ER-wave}) and (\ref{junction-tautau}) are
evolution equations and Eqs.(\ref{gammar}), (\ref{gammat}),
(\ref{explicit_intrinsic_junction}), (\ref{junction-psi}) and
(\ref{junction-r}) are constraint equations which the initial data
should be satisfied.
By the small modification of the static wall solutions in the previous
section, we obtain the momentarily static initial configurations that
satisfy these constraints.

%*************************************************************

%%%%%%%%%%%%%%%%%%%%%%%%%%%%%%%%%%%%%%%%%%%%%%%%%%%%%%%%%%%%%%%%%%%%%%
%\newpage
%%%%%%%%%%%%%%%%%%%%%%%%%%%%%%%%%%%%%%%%%%%%%%%%%%%%%%%%%%%%%%%%%%%%%%%
\subsection{Momentarily static initial configurations}
\label{sec:initial}
%%%%%%%%%%%%%%%%%%%%%%%%%%%%%%%%%%%%%%%%%%%%%%%%%%%%%%%%%%%%%%%%%%%%%%%

Let ${\cal S}$ be a momentarily static initial space and 
${\cal P} = {\cal S} \cap \Sigma$ is the initial locus of the wall on
${\cal S}$. 
Then ${\cal P}$ divides ${\cal S}$ into two parts: 
${\cal S}_{\pm}:={\cal S}\cap{\cal M}_{\pm}$.
We choose the origin of the comoving time $\tau=0$ at ${\cal P}$.

%*************************************************************

Further, we consider the radiation free initial condition, where there
is neither incidental nor outgoing ER waves on ${\cal S}$, i.e.,
$\partial_{t}^{2}\psi_{{\cal S}_{\pm}}(r) = \partial_{t}\psi_{{\cal
    S}_{\pm}}(r) = 0$. 
Then $\psi_{{\cal S}_{\pm}}(r)$ and $\gamma_{{\cal S}_{\pm}}(r)$ are
given by the same form as Eq.(\ref{waveless_solutions}).
In Eq.(\ref{waveless_solutions}), we denote $\gamma_{0}$ and $\kappa$
on ${\cal S}_{\pm}$ by $\gamma_{0\pm}$ and $\kappa_{\pm}$ as in the
last section.
We consider the situation where the wall is $r=R_{0}$ initially as
Eqs.(\ref{static_wall_condition}). 
By this choice, Eqs.(\ref{explicit_intrinsic_junction}) are trivially
satisfied.

%*************************************************************

Next, we evaluate the junction conditions
(\ref{junction-tautau})-(\ref{junction-r}). 
First, from (\ref{junction-psi}) and (\ref{junction-r}), we easily see
that $\kappa_{+}\neq\kappa_{-}$ by the regularity on ${\cal S}$. 
Actually, if $\kappa_{+}=\kappa_{-}$, (\ref{junction-psi}) and
(\ref{junction-r}), which have the same form as
Eqs.(\ref{waveless-junction-1}) and  (\ref{waveless-junction-2}),
shows that the axis $r=0$ is singular as seen in the last section. 
Further, the condition of Eqs.(\ref{waveless-junction-2})
together with the energy condition $\lambda>0$ and the $u^{t}_{\pm}>0$
leads $(\epsilon_{-},\epsilon_{+})=(+,\pm)$\cite{--case_com} and
${\cal S}_{-}$ includes the axis $r=0$. 
Then $\kappa_{+}\neq\kappa_{-}$. 
This means Eqs.(\ref{ur-utpm})-(\ref{epsilon_determination}) are also
true on ${\cal S}$.

%*************************************************************

In contrast to the results in Sec.\ref{sec:kappa+neqkappa-case},
the relation (\ref{Tomita-relation}), which was obtained from
Eq.(\ref{junction-tautau}), does not hold, since it contradicts to the
regularity of ${\cal S}$ as seen in the last section. 
Instead, the initial acceleration of the domain wall 
Eq.(\ref{acceleration_kappa-neqkappa+}) does not vanish. 
Then the domain wall begins to move and the system evolves to the
dynamical phase. 
The regular initial configurations are classified into two cases:
$(\epsilon_{-},\epsilon_{+})=(+,+),(+,-)$. 
In these cases, the motion of the wall is different. 
Then, we consider these cases, separately.

%%%%%%%%%%%%%%%%%%%%%%%%%%%%%%%%%%%%%%%%%%%%%%%%%%%%%%%%%%%%%%%%%%%%%%
%\newpage
%%%%%%%%%%%%%%%%%%%%%%%%%%%%%%%%%%%%%%%%%%%%%%%%%%%%%%%%%%%%%%%%%%%%%%%
\subsubsection{$\epsilon_{+} = 1$ case}
\label{sec:initial_case1}
%%%%%%%%%%%%%%%%%%%%%%%%%%%%%%%%%%%%%%%%%%%%%%%%%%%%%%%%%%%%%%%%%%%%%%%

To avoid the curvature singularity at $r=0$ in ${\cal S}_{-}$,
$\kappa_{-}$ should be $-1$ or $0$. 
In the following, we may concentrate only on the case $\kappa_{-}=0$
because $\kappa_{-}=-1$ case is locally equivalent to the case
$\kappa_{-}=0$. 
Actually, when $\kappa_{-}=-1$, $z$ should be a periodic coordinate
with the period $2\pi R_{0}e^{\gamma_{0-}}$ to avoid the conical
singularity at $r=0$ and the Killing orbit of $z^{a}$ on 
${\cal S}_{-}$ is not $R^{1}$ but $S^{1}$. 
Since $z$ is extended as a function on the whole spacetime ${\cal M}$,
$z$ is also a periodic coordinate on ${\cal S}_{+}$. 
We can easily check that this case is equivalent to the case
$\kappa_{-}=0$ with the above periodicity of $z$ by the following
replacements: $e^{\gamma_{0-}}t \rightarrow t$, $e^{\gamma_{0-}}r
\rightarrow r$,  $e^{-\gamma_{0-}}z/R_{0} \rightarrow \phi$,
$R_{0}\phi \rightarrow z$,  $-\kappa_{+}-1 \rightarrow \kappa_{+}$,
$e^{\gamma_{0-}}R_{0}\rightarrow R_{0}$ and $\gamma_{0+}-\gamma_{0-}
\rightarrow \gamma_{0+}$.

%*************************************************************

When $\kappa_{-}=0$, the conical singularity avoidance at $r=0$ in
${\cal S}_{-}$ leads $\gamma_{0-}=0$. 
Then, Eqs.(\ref{kpkm}) give 
\begin{eqnarray}
  \label{momentarily_static_X_special}
  && u^{t}_{+} = 1 - 2\lambda R_{0}, \quad u^{t}_{-} = 1,\\
  \label{kappa_p}
  && \kappa_{+} = \frac{\lambda R_{0}}{1 - 2 \lambda R_{0}}, \quad 
  e^{-\gamma_{0+}} = 1 - 2\lambda R_{0}.
\end{eqnarray}
Since $u^{a}$ is future directed ($u^{t}_{+}>0$),
Eqs.(\ref{momentarily_static_X_special}) lead
\begin{equation}
  \label{non_closed_radius}
  0 < R_{0} < \frac{1}{2\lambda},
\end{equation}
and $\kappa_{+}$ must be positive. The initial acceleration of the wall is
\begin{equation}
  \label{Minkow_kasner_accel}
  \left(\frac{d^{2}R}{d\tau^{2}}\right)_{0} = - \frac{2 - 3 \lambda
    R_{0}}{2 R_{0}}.
\end{equation}
By virtue of (\ref{non_closed_radius}), Eq.(\ref{Minkow_kasner_accel})
is the negative acceleration, i.e., the wall begins to collapse.

%*************************************************************

The geometry of the initial surface ${\cal S}$ is seen by evaluating
the relation between circumferential radius $\tilde{r}$ of the
symmetric cylinder defined by Eq.(\ref{circumference_phiphi}) and the
proper radial distance $\rho$ defined by 
\begin{equation}
  \label{proper_radial_distance}
  d\rho = e^{\gamma-\psi} dr = 
  e^{\gamma_{0+}} \left(\frac{r}{R_{0}}\right)^{\kappa_{+}^{2} +
    \kappa_{+}} dr,
\end{equation}
which is easily obtained by 
\begin{equation}
  \frac{d\tilde{r}}{d\rho} = \frac{1 + \kappa_{+}}{e^{\gamma_{0+}}}
  \left\{\frac{(\kappa_{+}^{2}+\kappa_{+}+1)\rho}{e^{\gamma_{0+}}R_{0}}
  \right\}^{\frac{-\kappa_{+}^{2}}{\kappa_{+}^{2}+\kappa_{+}+1}}.
\end{equation}
Since $\kappa_{+}>0$, $\tilde{r}(r)$ of the symmetric cylinder in
${\cal S}_{+}$ is a monotonically increasing function of $r$.
Then, ${\cal S}_{+}$ has infinity in the radial direction.
Further, in the limit $\lambda R_{0}\rightarrow 0$, (``mass per unit
proper $z$ length'' $\rightarrow 0$), we see $\kappa_{+}\rightarrow 0$
and $\gamma_{0+}\rightarrow0$.
It means ${\cal S}_{+}$ approaches to the flat space, i.e.,
$d\tilde{r}/d\rho \rightarrow 1$. 
We call this limit as ``{\it the weak gravity limit}''. 
This is the counter part of the test wall case. 
In the limit $\lambda R_{0} \rightarrow 1/2$, (equivalently ``mass per
unit proper $z$ length'' $\rightarrow 1/(4 G)$,) the initial
configuration characterized by Eqs.(\ref{kappa_p}) looks singular.
In this limit, $d\tilde{r}/d\rho \rightarrow 0$, which means the
circumference is constant outward. 
This curious geometry of ${\cal S}_{+}$ comes from the strong
gravitational effect of the wall. 
We call this limit as ``{\it the strong gravity limit}''. 
(See Appendix \ref{sec:closed_non_closed}.)

%*************************************************************

%%%%%%%%%%%%%%%%%%%%%%%%%%%%%%%%%%%%%%%%%%%%%%%%%%%%%%%%%%%%%%%%%%%%%%
%\newpage
%%%%%%%%%%%%%%%%%%%%%%%%%%%%%%%%%%%%%%%%%%%%%%%%%%%%%%%%%%%%%%%%%%%%%%%
\subsubsection{$\epsilon_{+} = -1$ case}
\label{sec:initial_case2}

In this case, both ${\cal S}_{\pm}$ contain the points $r=0$ and we
impose the regularity there. 
To avoid the curvature singularity, $\kappa_{-}$ should be $0$ or
$-1$. 
When $\kappa_{-}=0$, we choose $\gamma_{0-}=0$ to avoid the conical
singularity at $r=0$ on ${\cal S}_{-}$. 
Then ${\cal S}_{-}$ is flat. 
$\kappa_{+}\neq\kappa_{-}=0$ and the regularity at $r=0$ in ${\cal
  S}_{+}$ yields $\kappa_{+}=-1$. 
Eqs.(\ref{kpkm}) and (\ref{ur-utpm}) tell us $\lambda R_{0} = 1$ and
$\gamma_{+}=0$. 
Then ${\cal S}_{+}$ is also flat and the metric on ${\cal S}_{+}$ is given by
\begin{equation}
  \label{inner_metric_case2}
  ds^{2} = - dt^{2} + dr^{2} + \left(\frac{r}{R_{0}}\right)^{2} dz^{2} 
   + R_{0}^{2}d\phi^{2}. 
\end{equation}
From the regularity at $r=0$ in ${\cal S}_{+}$, the function $z$
should be a periodic coordinate with the period $2\pi R_{0}$ on
${\cal S}_{+}$. 
Since $z$ is extended so that the function on ${\cal M}$, $z$ should
be a periodic coordinate with the period $2\pi R_{0}$ also on 
${\cal S}_{-}$.  
Thus the initial surface ${\cal S}$ is closed and locally flat except
${\cal P}$. 
Further, we can easily see that the case $\kappa_{-}=-1$ is equivalent
to the case $\kappa_{-}=0$ by the replacement 
${\cal M}_{-}\leftrightarrow{\cal M_{+}}$.

%*************************************************************

The initial acceleration of the wall, in this case, is 
\begin{equation}
  \label{Minkow_kasner_accel_case3}
  \left(\frac{d^{2}R}{d\tau^{2}}\right)_{0} = \frac{\lambda}{2},
\end{equation}
i.e., the domain wall begins to expand.

%*************************************************************

Thus, domain walls on momentarily static, radiation free, and regular
initial configuration have the finite acceleration and do begin to
move in both case in the next moment. 
The case in Sec.\ref{sec:initial_case1} has the counter part in a test
wall system as seen in Appendix \ref{sec:test_wall}, while the case in
Sec.\ref{sec:initial_case2} does not. 
Henceforth, we concentrate on the case in Sec.\ref{sec:initial_case1}
to compare the self-gravitating wall with the test wall.

%%%%%%%%%%%%%%%%%%%%%%%%%%%%%%%%%%%%%%%%%%%%%%%%%%%%%%%%%%%%%%%%%%%%%%
%\newpage
%%%%%%%%%%%%%%%%%%%%%%%%%%%%%%%%%%%%%%%%%%%%%%%%%%%%%%%%%%%%%%%%%%%%%%%
\subsection{Einstein-Rosen wave emission}
\label{sec:emission}
%%%%%%%%%%%%%%%%%%%%%%%%%%%%%%%%%%%%%%%%%%%%%%%%%%%%%%%%%%%%%%%%%%%%%%%

\begin{figure}[h]
  \begin{center}
    \leavevmode
    \epsfxsize=0.40\textwidth
    \epsfbox{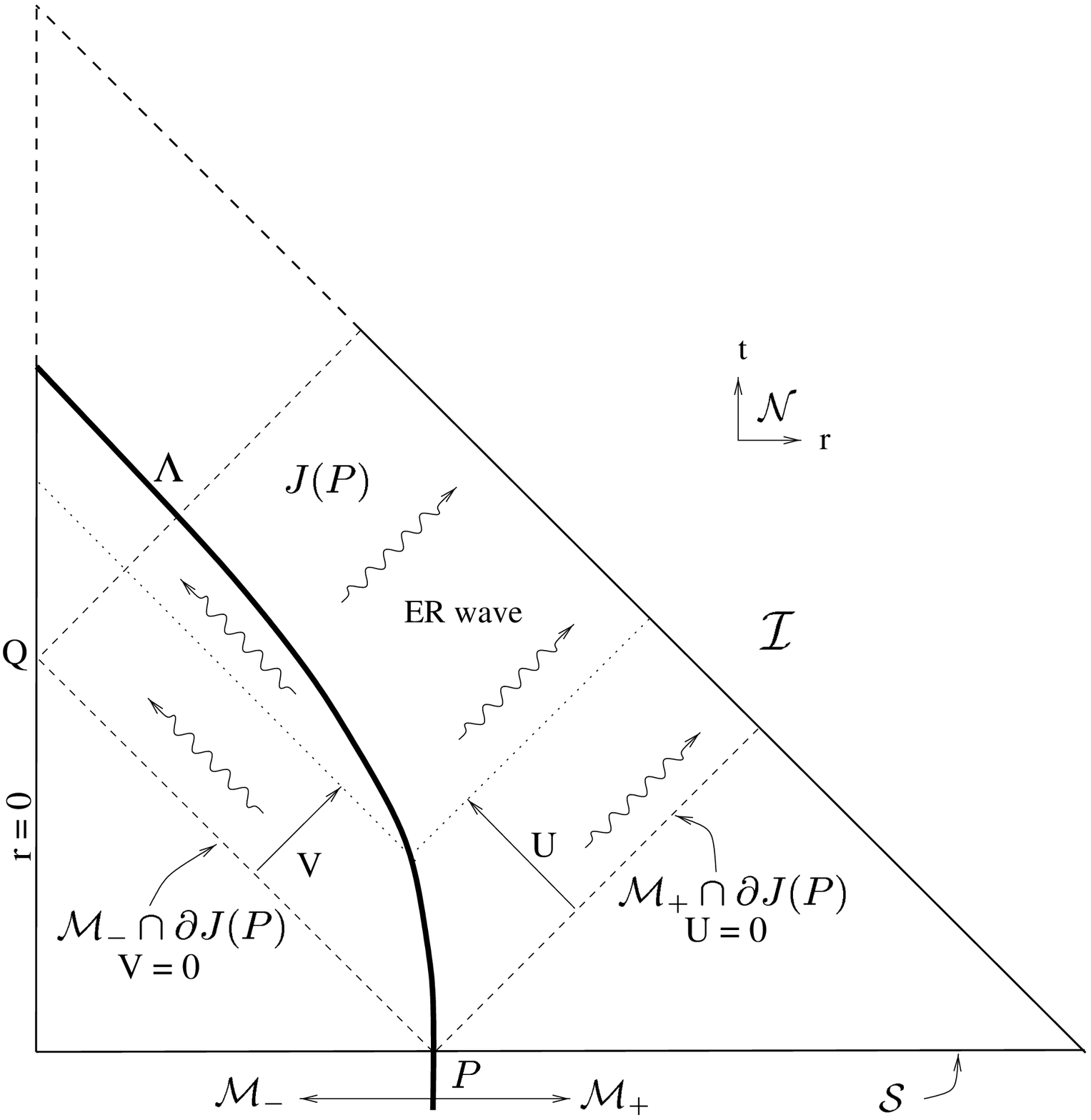}
    \caption{The schematic picture of the orbit space ${\cal N}$. 
      To clarify the time evolution from the initial surface ${\cal
      S}$, we should consider the causal future $J(P)$ of $P$ and
      $J({\cal S})\backslash J(P)$, separately.}
    \label{fig:cauchy}
  \end{center}
\end{figure}

Here, we consider the infinitesimal time evolution from the
momentarily static and radiation free initial configuration in
Sec.\ref{sec:initial_case1}. 
We separate the causal future $J({\cal S})$ of ${\cal S}$ in ${\cal
  M}$ into three pieces: the future domains of dependence 
$D^{+}({\cal S}_{\pm})$ of ${\cal S}_{\pm}$ and the causal future
$J({\cal P})$ of ${\cal P}$. 
(See Fig.\ref{fig:cauchy}.) 
We may treat these three pieces, separately. 
Since the initial configurations on $S_{\pm}$ are momentarily static
and radiation free, $D^{+}({\cal S}_{\pm})$ is still static and the
metric on $D^{+}({\cal S}_{\pm})$ is also given by
Eqs.(\ref{waveless_solutions}). 
Then, we may consider the time evolution from  $\partial J({\cal P})$.

%*************************************************************
%%%%%%%%%%%%%%%%%%%%% FIGURE 1. %%%%%%%%%%%%%%%%%%%%%%%%%%%%%%%%%
%%%%%%%%%%%%%%%%%%%%% FIGURE 1. %%%%%%%%%%%%%%%%%%%%%%%%%%%%%%%%%

To consider the evolution from $\partial J({\cal P})\cap{\cal M}_{+}$
into $J({\cal P})\cap {\cal M}_{+}$, we introduce the null coordinate
$U=t_{+}-r+R_{0}$ so that $U=0$ at $\partial J({\cal P})\cap{\cal M}_{+}$. 
In the coordinates $(U,r)$, the wave equation (\ref{ER-wave}) is given by 
\begin{equation}
  \label{ur_representation}
  \left(2\partial_{r} + \frac{1}{r}\right) \partial_{U}\psi -
  \left(\partial_{r}^{2} + \frac{1}{r}\partial_{r}\right)\psi = 0.
\end{equation}
Since we only consider the infinitesimal time evolution from 
$\partial J({\cal P})\cap{\cal M}_{+}$, we use the Taylor expansion of
solution $\psi = \psi_{+}$ around $\partial J({\cal P})\cap{\cal M}_{+}$ as 
follows:  
\begin{eqnarray}
  \label{Taylor}
  \psi_{+} &=& \frac{1}{\sqrt{r}}
  \sum_{n=0}^{\infty}\frac{1}{n!} \varphi^{(n)}_{+} (\omega U)^{n},  \\
  \label{outer_sol-1}
  \varphi^{(0)}_{+} &=& - \kappa_{+}\sqrt{r}\ln(r/R_{0}), \\
  \label{outer_sol-2}
  \psi^{(n)}_{+} &=& \sum^{n-1}_{l=0} a^{+}_{n-l}
  \frac{((2l-1)!!)^{2}}{l!} \left(\frac{-1}{8\omega r}\right)^{l} 
  \;\; \mbox{for} \;\; n>0,
\end{eqnarray}
where $a_{n}^{+}$ are constants determined by the junction condition
and $\omega^{-1}$ is an appropriate time scale. 
We have used the fact that the metric in the region $D^{+}({\cal
  S}_{+})$ is given by Eqs.(\ref{waveless_solutions}) to determine
$\varphi^{(0)}_{+}$.

%*************************************************************

Similarly, in ${\cal M}_{-}$, we expand $\psi$ by the advanced time
interval $V=t_{-}+r-R_{0}$ from $\partial J({\cal P})\cap{\cal M}_{-}$. 
The solution $\psi_{-}(V,r)$ is obtained as follows: 
\begin{eqnarray}
  \label{inner_solution}
  \psi_{-} &=& \frac{1}{\sqrt{r}}
  \sum_{n=0}^{\infty}\frac{1}{n!}\varphi^{(n)}_{-} (\omega V)^{n}, \\
  \label{inner_solution-2}
  \varphi^{(0)}_{-} &=& 0, \\
  \label{inner_solution-3}
  \varphi^{(n)}_{-} &=& \sum^{n-1}_{l=0} a^{-}_{n-l}
  \frac{((2l-1)!!)^{2}}{l!} \left(\frac{1}{8\omega r}\right)^{l},
  \; \mbox{for} \; n>0, 
\end{eqnarray}
where $a_{n}^{-}$ are the constants, which are determined by the
junction condition (\ref{junction-psi}).

%*************************************************************
%%%%%%%%%%%%%%%%%%%%%%%%%%%%%%%%%%%%%%%%%%%%%%%%%%%%%%%%%%%%%%%%%%%%%%
%\newpage
%%%%%%%%%%%%%%%%%%%%%%%%%%%%%%%%%%%%%%%%%%%%%%%%%%%%%%%%%%%%%%%%%%%%%%% 

Here, we consider the behavior of the solutions $\psi_{+}$ at
$r\rightarrow\infty$. 
Let ${\cal I}$ denotes the asymptotic region where
$r\rightarrow\infty$ with a fixed $U>0$. 
(See Fig.\ref{fig:cauchy}). 
The asymptotic form of $\psi_{+}$ at ${\cal I}$ is
\begin{equation}
  \label{Taylor-series-plus-partII}
  \psi_{+} = - \kappa_{+}\ln\frac{r}{R_{0}} +
  \frac{1}{\sqrt{r}} \sum^{\infty}_{n=1} \frac{a_{n}^{+}(\omega
    U)^{n}}{n!} + O\left(\frac{1}{r^{3/2}}\right). 
\end{equation}
The first term in Eq.(\ref{Taylor-series-plus-partII}) is the static
potential produced by the wall and the second term is the outgoing ER
wave. 
Eq.(\ref{Taylor-series-plus-partII}) shows $a_{n}^{+}$ correspond to
the amplitude of the ER wave at infinity ($r\rightarrow\infty$). 
Actually, the energy loss by the ER wave emission is estimated by the
quasi-local energy introduced by  Thorne\cite{Thone-old}: 
\begin{equation}
  E = \frac{\gamma}{4G},
\end{equation}
which called {\it ``C-energy''}. 
The energy loss is estimated by 
\begin{equation}
  \partial_{U}E_{\infty}(U) :=
  \lim_{r\rightarrow\infty}\partial_{U}E(r,U). 
\end{equation}
Using Eqs.(\ref{gammat}) and (\ref{Taylor-series-plus-partII}), we obtain 
\begin{equation}
  \label{energy-loss-general-form}
  \partial_{U}E_{\infty}(U) = 
  - \frac{1}{2G} \left( 
    \sum^{\infty}_{n=1}\frac{a_{n}^{+}\omega^{n}U^{n-1}}{(n-1)!}
  \right)^{2}. 
\end{equation} 
It shows that the C-energy must decrease due to the outgoing ER wave
($\partial_{U}E_{\infty}<0$).

%*************************************************************

On the other hand, $\psi_{-}$ given by
Eqs.(\ref{inner_solution})-(\ref{inner_solution-3}) has singular
behavior at the axis $r=0$. 
This singular behavior is removed if the superposition of the
reflecting waves at $r=0$ is taken into account. 
Let $Q$ be the point where the characteristics 
$\partial J({\cal P})\cap{\cal M}_{-}$ and the world line of the axis
$r=0$ intersects. 
(See Fig.\ref{fig:cauchy}.) 
The reflected waves emerge at the point $Q$ and the waves propagate
into the causal future $J(Q)$ of $Q$. 
Thus, the solution $\psi_{-}$ given by
Eqs.(\ref{inner_solution})-(\ref{inner_solution-3}) is applicable in
the region $(J({\cal P})\backslash J(Q))\cap{\cal M}_{-}$.

%*************************************************************
%%%%%%%%%%%%%%%%%%%%%%%%%%%%%%%%%%%%%%%%%%%%%%%%%%%%%%%%%%%%%%%%%%%%%%
%\newpage
%%%%%%%%%%%%%%%%%%%%%%%%%%%%%%%%%%%%%%%%%%%%%%%%%%%%%%%%%%%%%%%%%%%%%%% 

Now, we determine the coefficients $a_{n}^{\pm}$ by imposing the
junction conditions. 
The junction conditions (\ref{explicit_intrinsic_junction}) and
(\ref{junction-psi}) are equivalent to the set of equations at ${\cal P}$:  
\begin{equation}
  \label{an_determine}
  D_{\parallel}^{n+1}[D_{\perp}\psi] = 0, 
  \quad D_{\parallel}^{n}[\psi] = 0, \quad \mbox{for} \quad 
  n\geq 0.
\end{equation}
After some calculations (see Appendix \ref{sec:calculation}), we obtain 
\begin{eqnarray}
  \label{a1sol}
  a_{1}^{\pm} &=& 0, \\
  \label{a2sol}
  \omega^{2}a_{2}^{\pm} &=& \pm
  \frac{\kappa_{+}}{2(u^{t}_{\pm})^{2}\sqrt{R_{0}}}
  \left(\frac{d^{2}R}{d\tau^{2}}\right)_{0}, \\ 
  \omega^{3}a_{3}^{\pm} &=&
  \frac{\kappa_{+}}{2(u^{t}_{\pm})^3\sqrt{R_{0}}} \left\{ \pm
    \left(\frac{d^3 R}{d\tau^{3}}\right)_{0} -
    \left(\frac{d^{2}u^{t}_{+}}{d\tau^{2}}\right)_{0} \right.
  \nonumber \\  
  \label{a3sol}
  &+& \left.
    \frac{u^{t}_{+}}{R_{0}}\left(\frac{d^{2}R}{d\tau^{2}}\right)_{0}
  \right\} 
  \pm \frac{3\omega^{2}a_{2}^{\pm}}{(u^{t}_{\pm})^2}
  \left(\frac{d^2 R}{d\tau^{2}}\right)_{0} \\
  &\pm& \frac{\omega^{2}a_{2}^{\pm}}{8R_{0}} 
  - \frac{1}{4R_{0}(u^{t}_{\pm})^{3}}\left[ (u^{t}_{+})^{3}
    \omega^{2}a_{2}^{+} - (u^{t}_{-})^{3} \omega^{2}a_{2}^{-}
  \right], \nonumber
\end{eqnarray}
where all quantities in right hand side of
Eqs.(\ref{a1sol})-(\ref{a3sol}) are evaluated at ${\cal P}$ and
explicit form of $(d^{2}u^{t}_{+}/d\tau^{2})_{0}$ is given by
Eq.(\ref{utddot_final}) in Appendix \ref{sec:calculation}.

%*************************************************************

Eq.(\ref{a2sol}) shows that the ER wave emission is due to the wall
motion $(d^{2}R/d\tau^{2})_{0}$ in the leading order.
Eq.(\ref{a2sol}) includes two effects in ER wave emission. 
First, $\kappa_{+}$ plays a role of the source of ER wave. 
As seen in Appendix \ref{sec:calculation}, this comes from the
difference between $\partial_{r}\psi_{\pm}$ on ${\cal S}$. 
Second, the amplitude of the ER wave depends on $u^{t}_{+} =
e^{-\gamma_{+}}$, which represents the scale difference of the proper
time $\tau$ and the time coordinate $t_{+}$ (or $U$) on ${\cal S}_{+}$. 
These two effects caused by the self-gravity of the wall. 
Eq.(\ref{a3sol}) contains the influence due to the ER wave emission in
the next moment through Eq.(\ref{a2sol}).

%*************************************************************
%%%%%%%%%%%%%%%%%%%%%%%%%%%%%%%%%%%%%%%%%%%%%%%%%%%%%%%%%%%%%%%%%%%%%%
%\newpage
%%%%%%%%%%%%%%%%%%%%%%%%%%%%%%%%%%%%%%%%%%%%%%%%%%%%%%%%%%%%%%%%%%%%%%% 

Differentiating Eq.(\ref{junction-tautau}) with respect to $\tau$, we
obtain 
\begin{eqnarray}
  \label{dddr-1}
  \left(\frac{d^{3}R}{d\tau^3}\right)_{0} 
  &=& - \frac{2 \kappa_{+}(u^{t}_{+})^{3}}{(1 - u^{t}_{+})\sqrt{R_{0}}}
  \omega^{2} a_{2}^{+} \\
  \label{dddr-2}
  &=& \frac{\lambda (2 - 3 \lambda R_{0})}{4R_{0}(1 - 2\lambda
    R_{0})}.  
\end{eqnarray}
where we used Eqs.(\ref{derivative_coefficients}),
(\ref{momentarily_static_X_special}), (\ref{kappa_p}),
(\ref{Minkow_kasner_accel}) and (\ref{a2sol}). 
As seen in Appendix \ref{sec:test_wall}, $(d^{3}R/d\tau^{3})_{0}$
vanishes for the test wall in Minkowski spacetime. 
The appearance of $(d^{3}R/d\tau^{3})_{0}$, which depends on the
amplitude $a^{+}_{2}$ of outgoing ER wave, arises from the back
reaction of ER emission. 
Since $(d^{3}R/d\tau^{3})_{0}$ has the opposite sign to the
acceleration, the back reaction of ER wave emission diminishes the
absolute value of the wall acceleration in this order.

%*************************************************************

Let $\tau_{b}$ be the time scale where the back reaction becomes
efficient, which is estimated by
$|(d^{3}R/d\tau^{3})_{0}\tau_{b}|\sim|(d^{2}R/d\tau^{2})_{0}|$. 
From Eqs.(\ref{Minkow_kasner_accel}) and (\ref{dddr-2}), we found 
\begin{equation}
  \label{back_reaction_time}
  \tau_{b} \sim \frac{2(1 - 2 \lambda R_{0})}{\lambda}.
\end{equation}
In the strong gravity limit $\lambda R_{0} \rightarrow 1/2$,
$\tau_{b}$ shows that the radiation reaction becomes efficient
immediately. 
On the other hand, in the weak gravity limit $\lambda R_{0}\ll 1$,
Eqs.(\ref{Minkow_kasner_accel}) and (\ref{back_reaction_time}) are
given by 
\begin{equation}
  \label{accel_reactiontime_weak}
  \left(\frac{d^{2}R}{d\tau^{2}}\right)_{0} \sim - \frac{1}{R_{0}},
  \;\;\;\;\;\; \tau_{b} \sim \frac{2}{\lambda}.
\end{equation}
The initial acceleration in the weak gravity limit coincides with that
of the test wall (see Appendix \ref{sec:test_wall}). 
The collapsing time scale $\tau_{c}$ is estimated by
$|(d^{2}R/d\tau^2)|(\tau_{c})^{2}/2\sim R_{0}$. 
Then we obtain $\tau_{c}\sim R_{0} \ll 2/\lambda \sim \tau_{b}$ in the
same limit. 
Thus, in the weak gravity limit, the domain wall collapses to $R=0$
before the radiation reaction becomes efficient.

%*************************************************************
%%%%%%%%%%%%%%%%%%%%%%%%%%%%%%%%%%%%%%%%%%%%%%%%%%%%%%%%%%%%%%%%%%%%%%
%\newpage
%%%%%%%%%%%%%%%%%%%%%%%%%%%%%%%%%%%%%%%%%%%%%%%%%%%%%%%%%%%%%%%%%%%%%%% 

Substituting Eqs.(\ref{a2sol}) and (\ref{a3sol}) into
Eq.(\ref{energy-loss-general-form}), we estimate the energy loss rate
of the system by the ER wave emission: 
\begin{eqnarray}
  \label{mass-loss-rate}
  \partial_{U}E_{\infty} &\sim& - \frac{1}{2G}a_{2+}^{2}\omega^{4} U^2
  \left( 1 + \frac{\omega^{3}a^{+}_{3}}{\omega^{2}a^{+}_{2}}U \right),  \\
  \label{mass-loss-rate-2}
  &=& - \frac{1}{32G} \frac{\lambda^{2} (2 - 3\lambda R_{0})^{2}}{(1 -
    2\lambda R_{0})^{6}R_{0}} U^2 \times \\
  && \left( 1 
    - \frac{11 - 14 \lambda R_{0} - 28 (\lambda R_{0})^{2} 
      + 40 (\lambda R_{0})^{3}}{8R_{0}(1 - 2\lambda R_{0})^2}U \right).
  \nonumber
\end{eqnarray}
Eq.(\ref{mass-loss-rate-2}) shows that the larger energy are carried
by ER wave, as the initial total mass of the wall is larger in the
leading order. 
A cylindrical domain wall with large total mass produces strong
gravitational field and large energy density of the ER wave emitted.

%*************************************************************

In the weak gravity limit, Eq.(\ref{mass-loss-rate-2}) is given by 
\begin{equation}
  \label{weak_energy_loss}
  \partial_{U}E_{\infty} \sim - \frac{\lambda^{2} U^{2}}{8GR_{0}}
  \left(1 - \frac{11 U}{8R_{0}}\right).
\end{equation}
This shows that Eq.(\ref{weak_energy_loss}) is valid until
$U\sim8R_{0}/11$ in the weak gravity limit.

%%%%%%%%%%%%%%%%%%%%%%%%%%%%%%%%%%%%%%%%%%%%%%%%%%%%%%%%%%%%%%%%%%%%%%
%\newpage
%%%%%%%%%%%%%%%%%%%%%%%%%%%%%%%%%%%%%%%%%%%%%%%%%%%%%%%%%%%%%%%%%%%%%%
%%%%%%%%%%%% Section 4 ``Summary and Discussion'' %%%%%%%%%%%%%%%%%%%%
\section{Summary and Discussion}
\label{sec:summary}
%%%%%%%%%%%%%%%%%%%%%%%%%%%%%%%%%%%%%%%%%%%%%%%%%%%%%%%%%%%%%%%%%%%%%%

%*****************************************************************

In summary, we have considered the spacetime with a self-gravitating
cylindrical domain wall using the thin wall approximation. 
We considered two classes of solutions, separately, solutions with and
without gravitational waves emission. 
First, we found two subclasses of solutions without gravitational wave
emission: static wall solutions and dynamically moving wall
solutions. 
The spacetimes that are described by the solutions have singularities
inevitably. 
Next, we set up momentarily static and radiation free initial
configurations of the system. 
We found that $0<${\it ``wall's mass per unit proper
  length''}$<1/(4G)$ should hold initially for regular initial
configurations. 
In the exceptional case of {\it ``wall's mass per unit proper
  length''}$=1/(2G)$, we found that the initial surface is closed
radially. 
In these two regular initial configurations, the domain wall has
non-vanishing initial acceleration.

%*****************************************************************

We also considered the time evolution from the above initial data
within the infinitesimal time interval. 
By the accelerated motion of the wall, ER wave emission does occur. 
And the wave emission affects the wall motion in the next moment. 
This is just the radiation reaction problem. 
In contrast to the test wall motion in Minkowski spacetime, the
radiation reaction diminishes the acceleration of the wall. 
We also found that the amplitude of the emitted wave depends on the
gravitational potential produced by the wall on the initial surface. 
Further, the back reaction of the wave emission to the wall motion
depends both on the amplitude of the wave and on the gravitational
potential. 
As the result, a cylindrical domain wall with the large initial mass
produces the large gravitational potential. 
The wall motion in the large potential yields the large amount of the
ER wave. 
By this strong ER wave emission the large back reaction to the wall
motion does occur.

%*****************************************************************

Here, we compare the motion of the self-gravitating wall with that of
the test wall. 
When $\lambda R_{0} \ll 1$, the back reaction to the wall motion by ER
wave emission is negligible since the wall collapse to $r=0$ before
the back reaction becomes significant. 
Then, when $\lambda R_{0} \ll 1$, the motion of a self-gravitating
wall is well approximated by those of test walls. 
In the weak gravity limit, the energy loss rate by ER wave emission is
estimated from Eq.(\ref{weak_energy_loss}). 
Since the collapsing time scale is given by $U\sim R_{0}$, the energy
loss rate per length along $z$ direction is 
\begin{equation}
  \label{estimate_weak_energy_loss}
  \partial_{u}E_{\infty} \sim - 2 \pi^{2} G \sigma^{2} R_{0}. 
\end{equation}
Though Eq.(\ref{weak_energy_loss}) is not valid when $U\sim
8R_{0}/11$, we have extrapolated it. 
We note that the energy loss rate (\ref{estimate_weak_energy_loss})
does agree with that roughly estimated from the quadrupole formula for
gravitational wave emission. 
This is our main conclusion in this paper.

%*****************************************************************

The results obtained here are quite reasonable and these are expected
from the behavior of the test Nambu-Goto wall in the absence of their
self-gravity. 
However, the behaviors of gravitating Nambu-Goto membranes obtained in
Refs.\cite{KoIshiFuji,kouchan-string} are different from those
of test membranes.
We must note that the behavior similar to that in
Refs.\cite{KoIshiFuji,kouchan-string} is also obtained in the ER wave
scattering by the cylindrical domain wall in the static background as
seen in Appendix \ref{sec:scattering}, i.e., there is no solution
which corresponds to the spontaneous oscillation of the wall.
Further, we see that the cylindrical domain wall considered in
Appendix \ref{sec:scattering} is unstable. 
The unstable mode is not oscillatory mode. 
Though our analysis in Appendix \ref{sec:scattering} is restricted to
the perturbation of ER wave, we expect that the oscillatory behavior
of the wall, the oscillatory behavior is same as those in
Refs.\cite{KoIshiFuji,kouchan-string}.

%*****************************************************************

Together with the results from some models in
Refs.\cite{KoIshiFuji,kouchan-string} and in this paper, we conjecture
that the oscillatory motion of test walls fails to approximate that of
a self-gravitating wall, but monotonically collapsing motions of the
self-gravitating wall are well approximated by those of the test wall
if its energy density is sufficiently small.
Further, the energy of gravitational waves emitted by monotonic
motions of the walls is estimated by the quadruple formula.

%*****************************************************************

Of course, in the dynamics of domain walls in the realistic situation
in the early universe, the effective equation of state of an
oscillating wall may change, analogous to that of the wiggling cosmic
string\cite{Vilenkin_Shellard}. 
This is pointed out by Bonjour et.al\cite{Bonjour}. 
This change will be the effect to be taken into account when we
discuss the dynamics of domain walls in the realistic situation in the
early universe. 
However, we should emphasize that our discussion is concentrated on
the difference between the oscillatory behavior of a gravitating
Nambu-Goto membrane and a test Nambu-Goto membrane. 
At least, the oscillatory behavior of gravitating Nambu-Goto membranes
is quite different from that of test membranes, while monotonically
collapsing motion is well approximated by the test membranes. 
This suggests that the energy of gravitational waves estimated by the
oscillating test membranes might be incorrect.  
Though the full dynamics of gravitating Nambu-Goto membranes is not
clear yet, to estimate the energy of gravitational waves from defects,
we should clarify the dynamics of Nambu-Goto membranes at first. 
After that, the effect of the change in the equation of state should
be included. 
Though the gravitational waves might be emitted by this change, this
physical process is different from that discussed here.

%*****************************************************************

%%%%%%%%%%%%%%%%%%%%%%%%%%%%%%%%%%%%%%%%%%%%%%%%%%%%%%%%%%%%%%%%%%%%%%
%\newpage
%%%%%%%%%%%%%%%%%%%%%%%%%%%%%%%%%%%%%%%%%%%%%%%%%%%%%%%%%%%%%%%%%%%%%%

%%%%%%%%%%%%%%%%%%%%%%%%%%%%%%%%%%%%%%%%%%%%%%%%%%%%%%%%%%%%%%%%%%%%%%
\section*{Acknowledgments}
%%%%%%%%%%%%%%%%%%%%%%%%%%%%%%%%%%%%%%%%%%%%%%%%%%%%%%%%%%%%%%%%%%%%%%
The author would like to thank H.~Ishihara, T.~Okamura, A.~Ishibashi,
K.~Nakao and T.~Mishima for valuable comments and discussions. The
author also thanks to M.~Tachiki, M.~Date and M.~Omote for their
encouragement.

%%%%%%%%%%%%%%%%%%%%%%%%%%%%%%%%%%%%%%%%%%%%%%%%%%%%%%%%%%%%%%%%%%%%%%
\appendix
%%%%%%%%%%%%%%%%%%%%%%%%%%%%%%%%%%%%%%%%%%%%%%%%%%%%%%%%%%%%%%%%%%%%%%

%%%%%%%%%%%%%%%%%%%%%%%%%%%%%%%%%%%%%%%%%%%%%%%%%%%%%%%%%%%%%%%%%%%%%%
\section{Weyl canonical form and strong gravity limit}
\label{sec:closed_non_closed}
%%%%%%%%%%%%%%%%%%%%%%%%%%%%%%%%%%%%%%%%%%%%%%%%%%%%%%%%%%%%%%%%%%%%%%

As mentioned in the main text, the cylindrically symmetric spacetime
we considered here is characterized by the existence of two commutable
spacelike Killing vectors, which are both hypersurface orthogonal. 
In this Appendix, we explicitly see that the metric on this spacetime
is reduced to Eq.(\ref{Weyl-cano}) using the Einstein equation even if
the wall exists.

The metric on the spacetime which has two hypersurface orthogonal
Killing vectors $(\partial/\partial\phi)^{a}$ and $(\partial/\partial
z)^{a}$ is given by
\begin{equation}
  \label{general-metric}
  \begin{array}{rcl}
  ds^{2} &=& g_{ab}dx^{a}dx^{b} \\
  &=& e^{2\psi} dz^2 + e^{-2\psi}(\beta^{2}d\phi^{2} 
  + f_{ab}d\bar{x}^{a}d\bar{x}^{b}),
  \end{array}
\end{equation}
where $f_{ab}$ is the two dimensional Lorentzian metric
($f_{ab}(\partial/\partial z)^{a} = f_{ab}(\partial/\partial\phi)^{a}
= 0$), and $f_{ab}$, $\psi$ and $\beta$ depend only on the two
dimensional coordinates $\bar{x}^{a}$. 
$e^{\psi}$ and $e^{-\psi}\beta$ are the norms of the Killing vector
$z^{a}$ and $\phi^{a}$, respectively.

%*************************************************************

By choosing an appropriate null coordinates, $f_{ab}$ in
Eq.(\ref{general-metric}) is written by the conformal flat form
without loss of generality: 
\begin{equation}
  f_{ab}d\bar{x}^{a}d\bar{x}^{b} = - e^{2\bar{\gamma}} dx^{+}dx^{-}.
\end{equation}
When one treat the cylindrical vacuum spacetime, the function $\beta$
is constrained by one of the components of the Einstein equations,
$R^{z}_{z} + R^{\phi}_{\phi} = 0$ ($R^{a}_{b}$ is the Ricci
tensor), which yields
\begin{equation}
  \partial_{x^{+}}\partial_{x^{-}}\beta = 0.
\end{equation}
The general solution to this equation is
\begin{equation}
  \label{betasol}
  \beta = f_{1}(x^{+}) + f_{2}(x^{-}),
\end{equation}
where $f_{1}(x^{+})$ and $f_{2}(x^{-})$ are arbitrary functions of
$x^{+}$ and $x^{-}$, respectively. 
The gradient of $\beta$ determines whether we may choose $\beta$ as a
spatial coordinate or not. 
Actually, when $\partial_{a}\beta$ is spacelike, we may introduce
new null coordinates $\bar{u}$ and $\bar{v}$ so that
$\beta=(\bar{u}-\bar{v})/2$, and new coordinates $t$ and $r$ by
$t=(\bar{u}+\bar{v})/2$ and $r=(\bar{u}-\bar{v})/2$. 
Hence, one may perform the conformal transformation on the
$(x^{+},x^{-})$ plane to $(t,r)$ plane so that $\beta = r$. 
Thus, we have the Weyl canonical form (\ref{Weyl-cano}).

%*************************************************************

We must note that the signature of $\partial_{a}\beta$ is determined 
whether $\beta$ should be chosen by the spatial coordinate or not. 
When $\partial_{a}\beta$ is timelike or null, $\beta$ should be
regarded as the time or null coordinate, respectively. 
Further, we also note that the signature of $\partial_{a}\beta$ also
depends on the Einstein equation. 
It is not trivial whether above reduction is valid when the wall exists. 
Here, we show that we may choose $\beta$ as the radial coordinate
on the momentarily static initial surface when $\lambda R_{0}\neq
1/2$.

%*************************************************************

We only consider the case $\partial_{a}\beta$ is spacelike in 
${\cal M}_{-}$.
In the coordinate system (\ref{general-metric}), one of the junction
conditions is reduced to 
\begin{equation}
  \left[D_{\perp}\beta\right] = - 2 \lambda\beta.
  \label{eq:beta-junction}
\end{equation}
The condition (\ref{eq:beta-junction}) corresponds to
Eq.(\ref{junction-r}) in the main text.
Using the condition (\ref{eq:beta-junction}), we have
\begin{equation}
  \begin{array}{rcl}
    \left[ \partial_{a}\beta\partial^{a}\beta \right] &=& 
    \left[ - (D_{\parallel}\beta)^2 + (D_{\perp}\beta)^2 \right] \\
    &=& 
  - 4\left( (D_{\perp}\beta)_{-} - \lambda\beta \right) \lambda\beta,
  \end{array}
\end{equation}
and
\begin{equation}
  (\partial_{a}\beta\partial^{a}\beta)_{+} = -
  \left((D_{\parallel}\beta)_{-}\right)^{2} +
  \left((D_{\perp}\beta)_{-} - 2\lambda\beta\right)^{2}. 
\end{equation}
Since we choose $\beta=r$ on ${\cal S}\cap{\cal M}_{-}$, we have
$D_{\parallel}\beta=0$ and $D_{\perp}\beta = 1$ on ${\cal S}\cap{\cal M}_{-}$. 
Then we can see that $(\partial_{a}\beta)_{+}$ is spacelike when
$\lambda\beta \neq 1/2$.

%*************************************************************

When $\lambda\beta = 1/2$, $(\partial_{a}\beta)_{+}$ is null or
zero. 
Note that the case $\lambda\beta = 1/2$ corresponds to the ``strong
gravity limit'' in the main text. 
This geometry is obtained by taking limit $\kappa \rightarrow \infty$
in the metric (\ref{waveless_metric}). 
Using the proper radial coordinate defined by (\ref{proper_radial_distance}), we see that $g_{tt}$, $g_{zz}$ and $g_{\phi\phi}$ behave 
\begin{equation}
  g_{tt} \sim \kappa^{4} \left(\frac{\rho}{R_{0}}\right)^{2}, \quad
  g_{zz} \sim 1, \quad
  g_{\phi\phi} \sim R_{0}^{2}.
\end{equation}
in the limit $\kappa \rightarrow \infty$. Hence, the metric
(\ref{waveless_metric}) is given by  
\begin{equation}
  ds^{2} = - \rho^{2}dT^{2} + d\rho^{2} + dz^{2} + R_{0}^{2}d\phi^{2}. 
\end{equation}
This metric means ${\cal S}_{+}$ is locally flat but the circumference
of symmetric cylinder is constant $2\pi R_{0}$ outward as mentioned in
the main text. 
This metric shows that $\partial_{a}\beta$ is zero. 
This corresponds to the choice $\beta=R_{0}$, where $R_{0}$ is the
initial locus of the wall. 
This choice is required by the junction condition
(\ref{explicit_intrinsic_junction}).

%%%%%%%%%%%%%%%%%%%%%%%%%%%%%%%%%%%%%%%%%%%%%%%%%%%%%%%%%%%%%%%%%%%%%%
%\newpage
%%%%%%%%%%%%%%%%%%%%%%%%%%%%%%%%%%%%%%%%%%%%%%%%%%%%%%%%%%%%%%%%%%%%%%

%%%%%%%%%%%%%%%%%%%%%%%%%%%%%%%%%%%%%%%%%%%%%%%%%%%%%%%%%%%%%%%%%%%%%%
\section{Infinitesimal time evolution of ER wave}
\label{sec:calculation}
%%%%%%%%%%%%%%%%%%%%%%%%%%%%%%%%%%%%%%%%%%%%%%%%%%%%%%%%%%%%%%%%%%%%%%

In this appendix, we show the derivation of
(\ref{a1sol})-(\ref{a3sol}). 
Only we have to do is to evaluate the junction conditions
(\ref{explicit_intrinsic_junction}) and (\ref{junction-psi}) for
arbitrary $\tau$ and determine $a_{n\pm}$. 
Since we set up the momentarily static initial configuration, which
satisfy $\partial_{t}\psi =0$, i.e., $a_{1\pm}=0$. 
$a_{n}^{\pm}$ for $n\geq 1$ are determined by the evaluation of
(\ref{an_determine}) at ${\cal P}$. 
We consider the cases $(\epsilon_{-},\epsilon_{+}) = (+,\pm)$,
separately.

%%%%%%%%%%%%%%%%%%%%%%%%%%%%%%%%%%%%%%%%%%%%%%%%%%%%%%%%%%%%%%%%%%%%%%%
\subsection{$(\epsilon_{-},\epsilon_{+}) = (+,+)$ case}
%%%%%%%%%%%%%%%%%%%%%%%%%%%%%%%%%%%%%%%%%%%%%%%%%%%%%%%%%%%%%%%%%%%%%%%

To evaluate (\ref{an_determine}) for this case, it is convenient to
use the following representations of $D_{\para}$ and $D_{\perp}$; 
\begin{equation}
  \begin{array}{rcl}
    D_{\para+} &=& (u^{t}_{+} - u^{r})\partial_{U} + u^{r}\partial_{r}, \\
    D_{\perp+} &=& (u^{r} - u^{t}_{+})\partial_{U} + u^{t}\partial_{r}, \\
    D_{\para-} &=& (u^{t}_{-} + u^{r})\partial_{V} + u^{r}\partial_{r}, \\
    D_{\perp-} &=& (u^{r} + u^{t}_{-})\partial_{V} + u^{t}\partial_{r}.
  \end{array}
\end{equation}
Further, when we evaluates (\ref{an_determine}) at ${\cal P}$, the
momentarily static condition $u^{r}=\dot{u}^{t}_{\pm}=0$ are used. 
Note that $d^{n}u^{r}/d\tau^{n} \neq 0$ and
$d^{n+1}u^{t}_{\pm}/d\tau^{n+1} \neq 0$ for $n\geq2$. 
It is also convenient to use the expression
\begin{equation}
  \label{derivative_coefficients}
  \begin{array}{rcl}
    \partial_{r}^{n}\partial_{U}^{m}\psi_{{\cal P}+} &=& 
    \left.\omega^{m}\partial_{r}^{n}\left(
        \frac{\displaystyle \varphi_{+}^{(m)}}{\displaystyle \sqrt{r}} 
      \right)\right|_{r=R_{0}}, \\
    \partial_{r}^{n}\partial_{V}^{m}\psi_{{\cal P}-} &=& 
    \left.\omega^{m}\partial_{r}^{n}\left(
        \frac{\displaystyle \varphi_{-}^{(m)}}{\displaystyle \sqrt{r}} 
      \right)\right|_{r=R_{0}}.
  \end{array}
\end{equation}
Here, the subscript ``${\cal P}+$'' means the evaluation at $U=+0$ and
$r=R_{0}+0$ and ``${\cal P}-$'' means the evaluation at $V=+0$ and
$r=R_{0}-0$.

%*************************************************************

The junction conditions (\ref{an_determine}) with $n=2$ are given by 
\begin{eqnarray}
  && \omega^2 (u^{t}_{+})^{2}\varphi^{(2)}_{+} 
  + \omega^2 (u^{t}_{-})^{2}\varphi^{(2)}_{-} = 0, \\ 
  && \omega^2 (u^{t}_{+})^{2}\varphi^{(2)}_{+} 
  - \omega^2 (u^{t}_{-})^{2}\varphi^{(2)}_{-} \nonumber\\
  && \quad\quad =
  - \sqrt{r} \dot{u}^{r} \left.\left(
    \partial_{r}\left(\frac{\varphi^{(0)}_{+}}{\sqrt{r}} \right)
    - \partial_{r}\left(\frac{\varphi^{(0)}_{-}}{\sqrt{r}}
    \right) \right) \right|_{r=R_{0}}.
\end{eqnarray}
Then we obtain 
\begin{eqnarray}
  && \omega^2 (u^{t}_{\pm})^{2}\varphi^{(2)}_{\pm} \nonumber\\
  && = 
  \mp \frac{\sqrt{r}}{2} \dot{u}^{r} \left.\left(
    \partial_{r}\left(\frac{\varphi^{(0)}_{+}}{\sqrt{r}} \right)
    - \partial_{r}\left(\frac{\varphi^{(0)}_{-}}{\sqrt{r}}
    \right) \right) \right|_{r=R_{0}}.
  \label{varphi2_sol}
\end{eqnarray}
Since $\varphi^{(0)}_{\pm}/\sqrt{r} = - \kappa_{\pm}\ln(r/R_{0})$ and
$\kappa_{+}\neq\kappa_{-}$ from the regularity of the initial
configuration, $\varphi^{(2)}_{\pm}$ does not vanishes, i.e., the ER
wave emission occurs by the wall motion. 
When $\kappa_{-}=0$, Eq.(\ref{varphi2_sol}) gives Eq.(\ref{a2sol}) by
substituting $\varphi^{(0)}_{+} = - \kappa_{+} \sqrt{r}\ln(r/R_{0})$.

%*************************************************************

%%%%%%%%%%%%%%%%%%%%%%%%%%%%%%%%%%%%%%%%%%%%%%%%%%%%%%%%%%%%%%%%%%%%%%
%\newpage
%%%%%%%%%%%%%%%%%%%%%%%%%%%%%%%%%%%%%%%%%%%%%%%%%%%%%%%%%%%%%%%%%%%%%%
The conditions (\ref{an_determine}) with $n=3$ are given by  
\begin{eqnarray}
  && \omega^{3} (u^{t}_{+})^{3}\varphi^{(3)}_{+} + 
  \omega^{3} (u^{t}_{-})^{3}\varphi^{(3)}_{-} \nonumber\\
  &=& 
  3 \dot{u}^{r} \left( u^{t}_{+} \omega^{2} \varphi^{(2)}_{+} - 
    u^{t}_{-} \omega^{2} \varphi^{(2)}_{-} \right) \nonumber\\
  && + \sqrt{r} \dot{u}^{r} \left(
    u^{t}_{+} \partial_{r}^{2}\left(\frac{\varphi^{(0)}_{+}}{\sqrt{r}} \right)
    - u^{t}_{-} \partial_{r}^{2}\left(\frac{\varphi^{(0)}_{-}}{\sqrt{r}}
    \right) \right) \nonumber\\
  && + \sqrt{r} \left( (u^{t}_{+})^{3} \omega^{2}
    \partial_{r} \left(\frac{\varphi^{(2)}_{+}}{\sqrt{r}} \right) 
    - (u^{t}_{-})^{3} \omega^{2}
    \partial_{r} \left(\frac{\varphi^{(2)}_{-}}{\sqrt{r}}
    \right) \right) \nonumber\\
  \label{psi3minus}
  && \left. +  \sqrt{r} \left( 
    \ddot{u}^{t}_{+}\partial_{r}\left(\frac{\varphi_{+}^{(0)}}{\sqrt{r}}\right)
 - \ddot{u}^{t}_{-}\partial_{r}\left(\frac{\varphi_{-}^{(0)}}{\sqrt{r}}\right)
 \right)\right|_{r=R_{0}} , \\
  && (u^{t}_{+})^{3}\psi^{(3)}_{+} - (u^{t}_{-})^{3}\psi^{(3)}_{-} \nonumber\\
    &=&
   3 \dot{u}^{r} \left( u^{t}_{+} \omega^{2} \varphi^{(2)}_{+} + 
    u^{t}_{-} \omega^{2} \varphi^{(2)}_{-} \right) \nonumber\\
  \label{psi3plus}
  &&\left. - \sqrt{r} \ddot{u^{r}} \left(
    \partial_{r}\left(\frac{\varphi^{(0)}_{+}}{\sqrt{r}} \right)
    - \partial_{r}\left(\frac{\varphi^{(0)}_{-}}{\sqrt{r}}
    \right) \right) \right|_{r=R_{0}}. 
\end{eqnarray}
Then we obtain 
\begin{eqnarray}
  && \omega^{3} (u^{t}_{\pm})^{3} \varphi^{(3)}_{\pm} \nonumber\\
  &=& 
  \pm 3 \dot{u}^{r} u^{t}_{\pm} \omega^{2} \varphi_{\pm}^{(2)}
  \mp \frac{\sqrt{r}}{2} \ddot{u^{r}} 
  \left( \partial_{r}\left(\frac{\varphi^{(0)}_{+}}{\sqrt{r}} \right)
    - \partial_{r}\left(\frac{\varphi^{(0)}_{-}}{\sqrt{r}}
  \right) \right) \nonumber\\
  && + \frac{\sqrt{r}}{2} \dot{u}^{r} \left(
    u^{t}_{+} \partial_{r}^{2}\left(\frac{\varphi^{(0)}_{+}}{\sqrt{r}} \right)
    - u^{t}_{-} \partial_{r}^{2}\left(\frac{\varphi^{(0)}_{-}}{\sqrt{r}}
    \right) \right) \nonumber\\
  && + \frac{\sqrt{r}}{2} \left( (u^{t}_{+})^{3} \omega^{2}
    \partial_{r} \left(\frac{\varphi^{(2)}_{+}}{\sqrt{r}} \right) 
    - (u^{t}_{-})^{3} \omega^{2}
    \partial_{r} \left(\frac{\varphi^{(2)}_{-}}{\sqrt{r}}
    \right) \right) \nonumber\\
  && \left. +  \frac{\sqrt{r}}{2} \left( 
    \ddot{u}^{t}_{+}\partial_{r}\left(\frac{\varphi_{+}^{(0)}}{\sqrt{r}}\right)
 - \ddot{u}^{t}_{-}\partial_{r}\left(\frac{\varphi_{-}^{(0)}}{\sqrt{r}}\right)
 \right)\right|_{r=R_{0}} . 
  \label{varphi3_sol}
\end{eqnarray}
When $\kappa_{-}=0$, Eq.(\ref{varphi3_sol}) gives Eq.(\ref{a3sol}) by
substituting $\varphi^{(0)}_{+} = -\kappa_{+} \sqrt{r}\ln(r/R_{0})$.

%*************************************************************

Finally, $\ddot{u}^{t}_{+}$ in (\ref{psi3plus}) for the case in
Sec.\ref{sec:initial_case1} is derived as follows: Consider the norm
condition of $u^{a}$:
\begin{equation}
  \label{norm_cond}
  (u^{t}_{+})^{2} = (u^{r})^{2} + e^{-2(\Gamma_{+}-\Psi)}.
\end{equation}
Differentiating (\ref{norm_cond}) and using the momentarily static
conditions $u^{r} = \dot{\Psi} = \dot{\Gamma} = 0$ and $\Psi = 0$ at
${\cal P}$, we obtain 
\begin{equation}
  \label{utddot_org}
  u^{t}_{+}\ddot{u}^{t}_{+} = \ddot{R}^{2} - (\ddot{\Gamma}_{+} -
  \ddot{\Psi}) e^{- 2 \Gamma_{+}}.
\end{equation}
By virtue of (\ref{gammar}), (\ref{gammat}), (\ref{Taylor}) and
(\ref{derivative_coefficients}), we obtain 
\begin{eqnarray}
  \label{gamma_ddot_psi_ddot}
   \ddot{\Gamma}_{+} - \ddot{\Psi} 
   &=& \frac{\kappa_{+}(\kappa_{+} +
    1)}{R_{0}} \dot{u}^{r} \nonumber\\
  && - (2\kappa_{+} + 1) (u^{t}_{+})^{2}
  \frac{\omega^{2}}{\sqrt{R_{0}}}\varphi^{(2)}_{+}. 
\end{eqnarray}
Then (\ref{utddot_org}) gives
\begin{eqnarray}
  \label{utddot_final}
  \ddot{u^{t}_{+}}_{0} &=&
  \frac{1}{u^{t}_{+}} \left(\frac{d^2 R}{d\tau^{2}}\right)^{2}_{0} 
  - \frac{\kappa_{+}(\kappa_{+}+1)}{R_{0}} u^{t}_{+} \left(\frac{d^2
      R}{d\tau^{2}}\right)_{0} \nonumber\\
  && \quad\quad
+ (2\kappa_{+}+1) (u^{t}_{+})^3
  \frac{\omega^{2} a_{2}^{+}}{\sqrt{R_{0}}}.
\end{eqnarray}

%%%%%%%%%%%%%%%%%%%%%%%%%%%%%%%%%%%%%%%%%%%%%%%%%%%%%%%%%%%%%%%%%%%%%%%
\subsection{$(\epsilon_{-},\epsilon_{+}) = (+,-)$ case}
%%%%%%%%%%%%%%%%%%%%%%%%%%%%%%%%%%%%%%%%%%%%%%%%%%%%%%%%%%%%%%%%%%%%%%%

In this case, in both ${\cal M}_{\pm}$, ER wave propagates to the
direction along which the radial function $r$ decreases. 
Then, we should use  
\begin{eqnarray}
  \label{closed_outer_solution}
  \psi_{+} &=& \frac{1}{\sqrt{r}}
  \sum_{n=0}^{\infty}\frac{1}{n!}\varphi^{(n)}_{+} (\omega V_{+})^{n}, \\
  \label{closed_outer_solution-2}
  \varphi^{(0)}_{+} &=& \ln\left(\frac{r}{R_{0}}\right), \\
  \label{closed_outer_solution-3}
  \varphi^{(n)}_{+} &=& \sum^{n-1}_{l=0} a^{+}_{n-l}
  \frac{((2l-1)!!)^{2}}{l!} \left(\frac{1}{8\omega r}\right)^{l},
  \; \mbox{for} \; n>0, 
\end{eqnarray}
as the solution to (\ref{ER-wave}) in $J({\cal P})\cap {\cal M}_{+}$,
where $V_{+} = t_{+}+r-R_{0}$. 
Further, the representations of $D_{\para}$ and $D_{\perp}$ are given by 
\begin{eqnarray}
  \begin{array}{rcl}
    D_{\para+} &=& (u^{t}_{+} + u^{r})\partial_{V_{+}} + u^{r}\partial_{r}, \\
    D_{\perp+} &=& - (u^{r} + u^{t}_{+})\partial_{V_{+}} - u^{t}\partial_{r},\\
    D_{\para-} &=& (u^{t}_{-} + u^{r})\partial_{V_{-}} + u^{r}\partial_{r}, \\
    D_{\perp-} &=& (u^{r} + u^{t}_{-})\partial_{V_{-}} + u^{t}\partial_{r},
  \end{array}
\end{eqnarray}
where $V_{-} = V = t_{-}+r-R_{0}$.

The evaluations for (\ref{an_determine}) with $n=2$ by the similar
calculations to the case $(\epsilon_{-},\epsilon_{+})=(+,+)$ give the
same form as (\ref{varphi2_sol}). 
Since $\varphi^{(0)}_{+}/\sqrt{r} = \ln(r/R_{0})$ and
$\varphi^{(0)}_{-} = 0$, the ER wave emission occurs by the wall
motion. 
Furthermore, it is straight forward to obtain $\varphi^{(3)}_{\pm}$ by
the evaluation of (\ref{an_determine}) with $n=3$.

%*************************************************************

%%%%%%%%%%%%%%%%%%%%%%%%%%%%%%%%%%%%%%%%%%%%%%%%%%%%%%%%%%%%%%%%%%%%%%
\section{Dynamics of cylindrical test wall}
\label{sec:test_wall}
%%%%%%%%%%%%%%%%%%%%%%%%%%%%%%%%%%%%%%%%%%%%%%%%%%%%%%%%%%%%%%%%%%%%%%

We consider the test wall motion with the cylindrical symmetry for
comparison with the motion of the self-gravitating wall in the main text. 
The test wall motion is given by $K=0$, where $K$ is the trace of the
extrinsic curvature of the wall world volume. 
For a cylindrical test wall in Minkowski spacetime, the equation of
motion is given by
\begin{equation}
  \frac{d^{2}R}{d\tau^{2}} +
  \frac{1}{R}\left(\left(\frac{dR}{d\tau}\right)^{2} + 1\right)
  = 0.
\end{equation}
The solution to this equation is 
\begin{equation}
  R = \sqrt{R_{0}^{2} - \tau^{2}} \sim R_{0} -
  \frac{1}{2}\frac{\tau^{2}}{R_{0}} + R_{0}
  O\left(\left(\frac{\tau}{R_{0}}\right)^{4}\right).  
\end{equation}
The time symmetric initial surface is at $\tau=0$ and the acceleration
of the wall and the third derivative of $R$ at $\tau=0$ are given by
\begin{equation}
  \left(\frac{d^{2}R}{d\tau^{2}}\right)_{0} = - \frac{1}{R_{0}}, 
  \quad
  \left(\frac{d^{3}R}{d\tau^{3}}\right)_{0} = 0.
\end{equation}

%%%%%%%%%%%%%%%%%%%%%%%%%%%%%%%%%%%%%%%%%%%%%%%%%%%%%%%%%%%%%%%%%%%%%%
\section{Perturbative Einstein Rosen wave scattering by a
  Cylindrical domain wall} 
\label{sec:scattering}
%%%%%%%%%%%%%%%%%%%%%%%%%%%%%%%%%%%%%%%%%%%%%%%%%%%%%%%%%%%%%%%%%%%%%%

In this appendix, we consider the ER wave scattering by the domain
wall using one of the simplest static solutions obtained in
Sec.\ref{sec:waveless} and shows that the behavior of the wall is
similar to the case of the spherical wall\cite{KoIshiFuji}. 
This spacetime is an exact solution discussed by Ipser and
Sikivie\cite{Ipser-Sikivie}. 
The whole spacetime consists of two regions with the identical
Levi-Civita metric matched by the domain wall. 
On this background solution, we consider the perturbative ER wave and
the perturbative motion of the domain wall.

%*************************************************************

First, we describe the background spacetime and the ER wave
perturbation on this spacetime (Sec.\ref{sec:AppBG}). 
Second, we show the relation of the proper time $\tau$, the Gaussian
normal $\chi$ of the wall, and the coordinates $t$, $r$ in the main
text to solve the ER wave scattering by the domain wall
(Sec.\ref{sec:AppCoordinate}). 
Finally, we solve the scattering problem using the ``high frequency
approximation''(Sec.\ref{sec:AppScatSol}).

%*************************************************************

As mentioned in the main text, the background solution we use here has
curvature singularities. 
However, this singularity is not essential to our result.

%%%%%%%%%%%%%%%%%%%%%%%%%%%%%%%%%%%%%%%%%%%%%%%%%%%%%%%%%%%%%%%%%%%%%%
\subsection{Background and Perturbation} 
\label{sec:AppBG}
%%%%%%%%%%%%%%%%%%%%%%%%%%%%%%%%%%%%%%%%%%%%%%%%%%%%%%%%%%%%%%%%%%%%%%

As the background spacetime, we consider the spacetime in
Sec.\ref{sec:kappa+=kappa-case} with $\gamma_{0\pm}=0$. 
The line element is given by
\begin{equation}
  \label{Back-Ground}
  ds^{2} = \sqrt{\frac{R_{0}}{r}}(- dt^2 + dr^2) +
  \frac{r}{R_{0}}dz^2 + R_{0} r d\phi^2,
\end{equation}
The axis $r=0$ are curvature singularities and the Riemann polynomial
(\ref{riemann_polynomial}) diverge as $R^{abcd}R_{abcd} \propto
r^{-3}$ there and vanish at the infinity $r\rightarrow\infty$. 
Since the orbit space ${\cal N}$ is conformally flat, the conformal
diagram of ${\cal N}$ has timelike, spatial and null infinities.

%*************************************************************

\begin{figure}[h]
  \begin{center}
    \leavevmode
    \epsfxsize=0.25\textwidth
    \epsfbox[0 0 166 308]{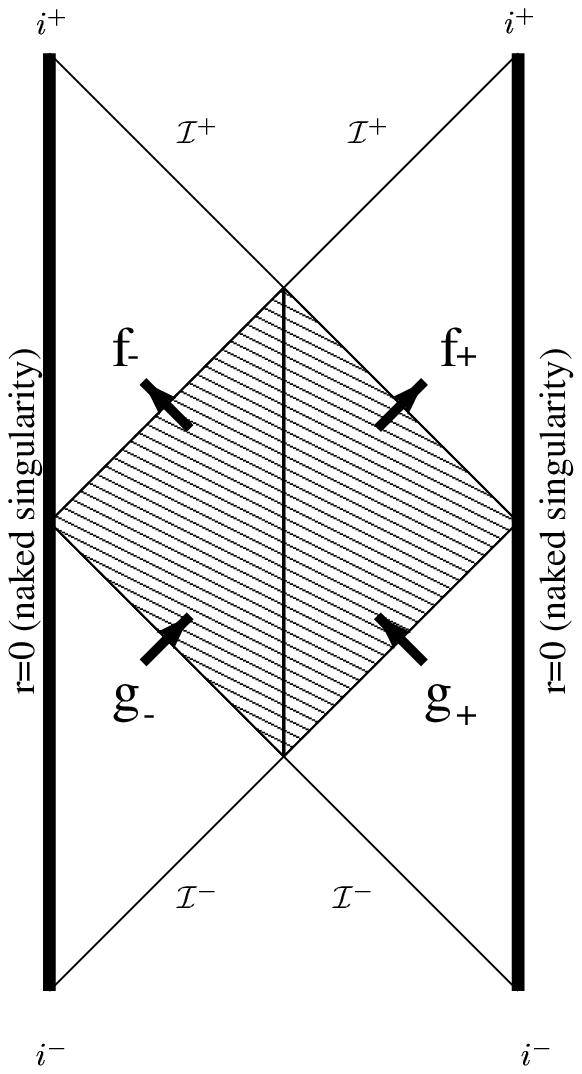} 
    \caption{The background spacetime ${\cal M}$ of our
    scattering problem. The shaded region is ${\cal D}$ defined
    by (\ref{calD-def}) and $\Sigma$ is contained in ${\cal D}$.}
    \label{fig:KK-junc}
  \end{center}
\end{figure}

In the spacetime in Sec.\ref{sec:kappa+=kappa-case} with
$\gamma_{0\pm}=0$, the motion of the wall is determined by the
equation (\ref{eq_of_motion_kappa+=kappa-}) and the trajectory
$\Sigma\cap{\cal N}$ on ${\cal N}$ is given by the equation: 
\begin{equation}
  \label{traofwall}
  \left(\frac{dr}{dt}\right)^{2} +
  \frac{1}{\lambda^{2}R_{0}^{1/2}r^{3/2}} - 1 = 0.
\end{equation}
The qualitative behavior of the wall motion is as follows: Seeing from
an observer whose world line is $r=$constant in either vacuum region,
the domain wall first starts to shrink from the past null infinity at
the light velocity, decelerates its speed, and bounces at a finite
radius 
\begin{equation}
  \label{turning}
  r_{0}:=\frac{1}{\lambda^{4/3}R_{0}^{1/3}}.   
\end{equation}
After that it expands to the null infinity. 
Then the solution to (\ref{traofwall}) is contained in the region 
\begin{equation}
  \label{calD-def}
  {\cal D} := \{ (t,r) | \quad r^2-t^2>0\quad\}
\end{equation}
in ${\cal N}$. 
The resulting spacetime ${\cal M}$ has the future and the past null
infinities each of which has two connected components. 
The global structure of this spacetime is analogous to that in
Ref.\cite{KoIshiFuji} but there are singularities at $r=0$. 
(See Fig.\ref{fig:KK-junc}.)

%*************************************************************

On this spacetime, we consider the perturbative deformation of the
wall due to the scattering of the perturbative ER wave:
\begin{equation}
  \label{psi0pluspsi1}
  \psi = \frac{1}{2} \ln\frac{r}{R_{0}} + \epsilon\varphi, \quad
  \gamma = \frac{1}{4} \ln \frac{r}{R_{0}} + \epsilon \varphi,
\end{equation}
where $\epsilon$ is the infinitesimal parameter for the perturbation. 
Since the equation (\ref{ER-wave}) is linear, $\varphi$ is governed by
the same equation as Eq.(\ref{ER-wave}). 
Here we have ignored the constant term in $\gamma$ which corresponds
to the additional deficit angle around axis $r=0$ due to the
perturbation.

%*************************************************************

To construct the global solution to $\varphi$, we must consider the
Israel's junction condition at $\Sigma$ for $\varphi$: 
\begin{equation}
  \label{wave-junction}
  [\varphi] = 0,\;\;   [D_{\perp}\varphi] = 0.
\end{equation}
We must also note that the junction condition (\ref{junction-r}) leads
to the same equation of motion of the domain wall as (\ref{traofwall})
in the order of $O(\epsilon)$ even if we include the perturbative term
$\epsilon\varphi$. 
However, the wall does undergo the quadrupole deformation by the
perturbed ER wave. 
Actually, the circumferential radius of the domain wall along the
Killing direction $\phi^{a}$ is fluctuated as
\begin{equation}
  {\cal R} = 2 \pi \sqrt{R_{0} R}(1 - \epsilon \varphi_{s})  
\end{equation}
within the linear order of $\epsilon$, where
$\varphi_{s}=\left.\varphi\right|_{\Sigma}$.
Further, the proper length along the Killing orbit of $z^{a}$ is
deformed as
\begin{equation}
 e^{\psi}dz =
 \left(\sqrt{\frac{r}{R_{0}}}+\epsilon\varphi_{s}\right)dz.
\end{equation}
Then, ER wave is directly related to the fluctuations of $\Sigma$ and
the intrinsic geometries of $\Sigma$ are fluctuated by the
perturbative ER wave $\varphi$ if $\varphi_{s}$ does not vanish. 
This situation is the same as that in Ref.\cite{KoIshiFuji}.

%*************************************************************

%%%%%%%%%%%%%%%%%%%%%%%%%%%%%%%%%%%%%%%%%%%%%%%%%%%%%%%%%%%%%%%%%%%%%%
\subsection{Coordinates on ${\cal D}$} 
\label{sec:AppCoordinate}
%%%%%%%%%%%%%%%%%%%%%%%%%%%%%%%%%%%%%%%%%%%%%%%%%%%%%%%%%%%%%%%%%%%%%%

To solve the ER wave scattering by the domain wall, it is convenient
to give the explicit relation between the coordinate system $(t,r)$ in 
the Weyl canonical form (\ref{Weyl-cano}) and the comoving coordinate
$(\tau,\chi)$ of the wall trajectory $\Sigma\cap{\cal N}$.

%*************************************************************

Suppose that the solution to Eq.(\ref{traofwall}) is given by 
\begin{equation}
  v = p(u)
\end{equation}
where $v=t+r$ and $u=t-r$. The asymptotic behavior of $p(u)$ is as follows:
\begin{eqnarray}
  \label{trajectory_behavior}
  p(u) \sim \left\{
    \begin{array}{ll}
      \frac{1}{\lambda^2\sqrt{R_{0}}}\sqrt{\frac{2}{-u}}
      & (u \rightarrow -\infty); \\ 
      \frac{2}{\lambda^{4}R_{0} u^2} & 
      (u \rightarrow -0).
    \end{array}
  \right.
\end{eqnarray}
Since $\Sigma\cap{\cal N}$ is timelike, 
$\Sigma\cap{\cal N}\subset{\cal D}$, where 
\begin{equation}
  {\cal D} = \{(v,u)|0<v<+\infty,-\infty<u<0\}.
\end{equation}

%*************************************************************

Here, we introduce new double null coordinates
\begin{equation}
  \label{x-uv-trans}
  q(\sigma^{+}) = v, \quad 
  q(\sigma^{-}) = p(u).
\end{equation}
$q$ is a monotonically increasing function determined so that the unit
normal $n^{a}$ of $\Sigma\cap{\cal N}$ is 
\begin{equation}
  \label{normal_null}
  n^{a} = \left(\frac{\partial}{\partial\sigma^{+}}\right)^{a} -
    \left(\frac{\partial}{\partial\sigma^{-}}\right)^{a}, 
\end{equation}
at least on $\Sigma\cap{\cal N}$. 
In terms of $\sigma^{\pm}$, $\Sigma\cap{\cal N}$ is given by
$\sigma^{+}=\sigma^{-}$ and the metric on $\cal D$ is given by  
\begin{equation}
  ds^{2}|_{\cal N} = - 2g(\sigma^{+},\sigma^{-}) d\sigma^{+}d\sigma^{-}, 
\end{equation}
where
\begin{eqnarray}
  g(\sigma^{+},\sigma^{-}) &=&
  \frac{1}{2}\sqrt{\frac{2R_{0}}{q(\sigma^{+}) - 
      p^{-1}\circ q(\sigma^{-})}}\times \nonumber\\
  && \quad \frac{1}{p'\circ q(\sigma^{-})}
  \frac{dq}{d\sigma^{+}}(\sigma^{+})
  \frac{dq}{d\sigma^{-}}(\sigma^{-}).
  \label{eq:g-explicit-form}
\end{eqnarray}
In (\ref{eq:g-explicit-form}), $p^{-1}(v)$ is the inverse function of
$v = p(u)$, $\circ$ denotes the composite function, and
$p'(v)=dp/du(u=p^{-1}(v))=dp/du \circ p^{-1}(v)$. 
Then, the monotonically increasing function $q$ is determined by
$n^{a}n_{a} = 1$ on $\Sigma\cap{\cal N}$ ($\sigma^{+}=\sigma^{-}$): 
\begin{equation}
  \label{q_sigma_plus_relation}
  \frac{dq}{d\sigma^{+}} = \sqrt{p'\circ
    q(\sigma^{+})\sqrt{\frac{q(\sigma^{+})-p^{-1}\circ
        q(\sigma^{+})}{2R_{0}}}}.
\end{equation}
Then, the coordinate transformation from $(u,v)$ to
$(\sigma^{-},\sigma^{+})$ is given by  
\begin{eqnarray}
  \label{relations-1}
  \sigma^{-} &=& \int du \sqrt{p'(u)\sqrt{\frac{2R_{0}}{p(u)-u}}},\\
  \label{relations-2}
  \sigma^{+} &=& \int dv \sqrt{\frac{1}{p'\circ
      p^{-1}(v)}\sqrt{\frac{2R_{0}}{v-p^{-1}(v)}}} 
\end{eqnarray}
from Eqs.(\ref{x-uv-trans}).

%*************************************************************

From Eq.(\ref{trajectory_behavior}), the asymptotic behaviors of the
coordinates $\sigma^{\pm}$ are 
\begin{eqnarray}
  \sigma^{-} \sim - \frac{1}{\lambda}\ln(-u), \quad
  \sigma^{+} \sim \frac{2}{\lambda}\ln v,\quad
  (u \rightarrow -\infty), \\
  \sigma^{-} \sim - \frac{1}{\lambda}\ln(-u), \quad
  \sigma^{+} \sim \frac{2}{\lambda}\ln v,\quad
  (u \rightarrow -0).
\end{eqnarray}
Then, $\sigma^{\pm}$ is $C^{1}$ function on the region $\cal D$ and
$\cal D$ is covered by the coordinate system: 
\begin{equation}
  {\cal D} = \{(\sigma^{+},\sigma^{-}) | -\infty<\sigma^{\pm}<\infty\}.
\end{equation}

%%%%%%%%%%%%%%%%%%%%%%%%%%%%%%%%%%%%%%%%%%%%%%%%%%%%%%%%%%%%%%%%%%%%%%
\subsection{ER wave scattering} 
\label{sec:AppScatSol}
%%%%%%%%%%%%%%%%%%%%%%%%%%%%%%%%%%%%%%%%%%%%%%%%%%%%%%%%%%%%%%%%%%%%%%

Using above coordinates $\sigma^{\pm}$ in ${\cal D}$, we consider the
scattering problem of ER wave by the cylindrical domain wall.
To solve the problem, we consider the ``high frequency approximation''
in the region $\cal D$.

%*************************************************************

In terms of the null coordinates $\sigma^{\pm}$, Eq.(\ref{ER-wave}) is 
given by  
\begin{equation}
  \label{ER-wave-perturb-2}
  \left(- \partial_{+}\partial_{-} +
    \frac{1}{16r(\sigma^{+},\sigma^{-})^2} \frac{dv}{d\sigma^{+}}
    \frac{du}{d\sigma^{-}} \right)(\sqrt{r}\varphi) = 0,
\end{equation}
where $\partial_{\pm}= \partial/\partial\sigma^{\pm}$. 
From the above construction of the coordinate system $\sigma^{\pm}$,
the proper time $\tau$ and the Gaussian normal coordinate $\chi$ of
$\Sigma\cap{\cal D}$($=\Sigma\cap{\cal N}$) are given by $\tau =
(\sigma^{+}+\sigma^{-})/2$, $\chi = (\sigma^{+}-\sigma^{-})/2$. 
Then, Eq.(\ref{ER-wave-perturb-2}) is also given by 
\begin{eqnarray}
  \label{ER-wave-perturb-3}
  && \left(- \partial_{\tau}^{2} + \partial_{\chi}^{2} + V(\tau,\chi)
  \right)(\sqrt{r}\varphi) = 0, \\
  \label{potential_function}
  && V(\tau,\chi) \equiv \frac{1}{4r(\sigma^{+},\sigma^{-})^2}
 \frac{dv}{d\sigma^{+}} \frac{du}{d\sigma^{-}}.
\end{eqnarray}

%*************************************************************

The ``high frequency approximation'' considered here is as follows:
Consider the Fourier decomposition of $\varphi = \int d\omega
e^{i\omega\tau}\varphi_{\omega}(\chi)$. 
Further, we denote the maximum of the absolute value of the potential
$V(\tau,\chi)$ along $\Sigma\cap{\cal D}$ by $||V(\tau,0)||$. 
If $\omega^{2}$ is sufficiently large, $\omega^{2} \gg ||V(\tau,0)||$,
we may ignore the potential term in Eq.(\ref{ER-wave-perturb-3}). 
From Eqs.(\ref{relations-1}) and (\ref{relations-2}), we obtain 
\begin{equation}
  \frac{du}{d\sigma^{-}} \frac{dv}{d\sigma^{+}} =
  \sqrt{\frac{r}{R_{0}}} 
\end{equation}
along the trajectory $\Sigma\cap{\cal D}$ and 
\begin{equation}
  ||V(\tau,0)|| = \frac{\lambda^2}{4}.
\end{equation}
Here, we note that $V(\tau,\chi)$ is maximum at the turning point
(\ref{turning}). 
Thus, the ``high frequency approximation'' here corresponds to the
concentration on the waves with the frequency much larger than the
wall tension.

%*************************************************************

In the ``high frequency approximation'', the solution to the wave
equation (\ref{ER-wave-perturb-3}) is given by
\begin{equation}
  \label{high_frequency_solution}
  \varphi_{\pm} \sim f(\sigma^{+})_{\pm} + g(\sigma^{-})_{\pm},
\end{equation}
where $f$ and $g$ are arbitrary functions which satisfy the conditions
\begin{equation}
  \label{approximation_condition}
 \partial^{2}_{+}f \gg \frac{\lambda^{2}}{4}f, \quad 
 \partial^{2}_{-}g \gg \frac{\lambda^{2}}{4}g.
\end{equation}
Evaluating the junctions conditions (\ref{wave-junction}) on
$\Sigma\cap{\cal D}$, we obtain
\begin{eqnarray}
  f_{\pm}(\sigma^{+})&& = c_{\pm} e^{\frac{\lambda\sigma^{+}}{2}} \nonumber\\
    + &&e^{\frac{\lambda\sigma^{+}}{2}}\int^{\sigma^{+}}d\hat{\tau}
    \left(\frac{dg_{\mp}}{d\tau} + \frac{\lambda}{2}
      g_{\pm}\right) e^{-\frac{\lambda\hat{\tau}}{2}}, 
  \label{scattering_data}
\end{eqnarray}
where $g_{\pm}$ and $f_{\pm}$ are incident and scattered wave,
respectively (See Fig.\ref{fig:KK-junc}), and $c_{\pm}$ are constants
of integration. 
We note that the first term does not depend on the incident waves,
while the second term of Eq.(\ref{scattering_data}) corresponds to the
reflected or transmitted waves of the incident waves. 
The solution 
\begin{equation}
  \label{non-trivial-scattering}
  f_{\pm}(\sigma^{+})=c_{\pm}e^{\lambda\sigma^{+}/2}  
\end{equation}
is singular at the future boundary of the region $\cal D$, which
does not satisfy the condition (\ref{approximation_condition}). 
Then, within our approximation, we conclude that the domain wall does
not emit gravitational waves spontaneously by its free oscillations in
the high frequency approximation. 
This conclusion is same as those in Ref.\cite{KoIshiFuji}.

%*************************************************************

We must note that the result obtained here does not depend on the
boundary condition on the singularity $r=0$ of the background
spacetime. 
Our consideration is restricted in the region ${\cal D}$ and the
singularity $r=0$ is not in ${\cal D}$ but on the boundary
$\partial{\cal D}$ of the closure of ${\cal D}$. 
From the causality determined by Eq.(\ref{ER-wave}), the singularity
$r=0$ on $\partial{\cal D}$ does not give any effects in ${\cal D}$.

%*************************************************************

We must also note the fact that the ``high frequency approximation''
discussed here gives exact scattering data in the model of
Ref.\cite{KoIshiFuji} and the solution similar to
Eq.(\ref{non-trivial-scattering}) is obtained. 
The solution obtained by this approximation is just $k=-i$ mode in
Ref.\cite{KoIshiFuji}.  
Then we may say that the solution based on the high frequency
approximation is also correct beyond the validity of the approximation
in the exactly solvable model in Ref.\cite{KoIshiFuji}. 
If we apply this extrapolation to the solution
(\ref{scattering_data}), the only solution in the absence of
incidental wave is (\ref{non-trivial-scattering}) and we conclude that
the background cylindrical domain wall considered here is unstable.

%%%%%%%%%%%%%%%%%%%% references %%%%%%%%%%%%%%%%%%%%%%%%%%%%%%%%%%%

\end{multicols}


\begin{thebibliography}{99}
\bibitem{notation-subst}
  We denote $[A] := A_{+} - A_{-}$ for arbitrary functions
  $A_{\pm}$ on $\Sigma_{\pm}$, respectively.
\bibitem{Ipser-Sikivie-comment}
  For a special case $\gamma_{0\pm}=0$ the solution is obtained
  in Ref.\cite{Ipser-Sikivie}
\end{thebibliography}
\end{document}